\documentclass[a4paper,fleqn,usenatbib]{mnras}
\usepackage[T1]{fontenc}
\usepackage{ae,aecompl}

\usepackage{mathptmx}
\usepackage{lscape}
\usepackage{url}
\usepackage{graphicx}
\usepackage{longtable}
\newcommand{\ts}{\thinspace}
\newcommand{\simless}{\mathbin{\lower 3pt\hbox
     {$\rlap{\raise 5pt\hbox{$\char'074$}}\mathchar"7218$}}}
\newcommand{\simgreat}{\mathbin{\lower 3pt\hbox
     {$\rlap{\raise 5pt\hbox{$\char'076$}}\mathchar"7218$}}}
\newcommand{\msun}{\ts M$_\odot$}

\newcommand{\teff}{T$_{\rm eff}$}
\newcommand{\logt}{$\log({\rm T_{eff}})$}
\newcommand{\logl}{$\log({\rm L/L_{\odot}})$}

\newcommand{\angstrom}{\textrm{\AA}}
\newcommand{\masyr}{\ts mas~yr$^{-1}$}

\title[K-Type Members of Sco-Cen]{The Star-formation History and Accretion-Disk Fraction Among the K-Type
Members of the Scorpius-Centaurus OB Association}

\author[Pecaut \& Mamajek]{
Mark~J.~Pecaut,$^1$\thanks{E-mail: mark.pecaut@rockhurst.edu}
Eric~E.~Mamajek$^2$ 
\\
$^{1}$Rockhurst University, Department of Physics, 1100 Rockhurst Rd, Kansas City, MO 64110-2508, USA \\
$^{2}$University of Rochester, Department of Physics and Astronomy, Rochester, NY 14627-0171, USA
}

\date{Accepted XXX. Received YYY; in original form ZZZ}

\pubyear{2015}

\begin{document}
\label{firstpage}
\pagerange{\pageref{firstpage}--\pageref{lastpage}}
\maketitle

\begin{abstract}
%
We present results of a spectroscopic survey for new K- and M-type
members of Scorpius-Centaurus (Sco-Cen), the nearest OB Association
($\sim$100--200 pc).
Using an X-ray, proper motion and color-magnitude selected sample, we
obtained spectra for 361 stars, for which we report spectral
classifications and Li and H$\alpha$ equivalent widths.
We identified 156 new members of Sco-Cen, and recovered 51 previously
published members.  We have combined these with previously known members
to form a sample of 493 solar-mass ($\sim$0.7--1.3 \msun) members of Sco-Cen.
We investigated the star-formation history of this sample, and
re-assessed the ages of the massive main-sequence turn-off and
the G-type members in all three subgroups.
We performed a census for circumstellar disks in our sample using
{\it WISE} infrared data and find a protoplanetary disk fraction for K-type
stars of 4.4$^{+1.6}_{-0.9}$\% for Upper Centaurus-Lupus and Lower 
Centaurus-Crux at $\sim$16~Myr and 9.0$^{+4.0}_{-2.2}$\% for Upper Scorpius at
$\sim$10 Myr.
These data are consistent with a protoplanetary disk e-folding
timescale of $\sim$4--5 Myr for $\sim$1~\msun\, stars, twice that previously
quoted (Mamajek 2009), but consistent with the Bell et al. revised age scale 
of young clusters.  
Finally, we construct an age map of Scorpius-Centaurus which 
clearly reveals substructure consisting of concentrations of younger and older 
stars.  We find evidence for strong age gradients within all three subgroups.  
None of the subgroups are consistent with being simple, coeval populations
which formed in single bursts, but likely represents a multitude of
smaller star formation episodes of hundreds to tens of stars each.
\end{abstract}





\section{Introduction and Background}

Because of their utility for studying star and planet formation and
evolution, a great deal of effort has been directed towards
identifying and characterizing samples of nearby young stars.  These
samples have been indispensible for direct imaging of brown dwarf and 
planetary-mass companions \citep[e.g.,][]{kraus2012}, imaging debris disks
in scattered light \citep[e.g.,][]{kalas2015}, exploring the
star-disk interaction in young stars \citep[e.g.,][]{nguyen2009}, and
studying the evolution of gas-rich and dusty debris disks as a
function of age \citep[e.g.,][]{mamajek2009,chen2011}.

The nearest of these samples are small, diffuse, nearby associations
of kinematically related stars, (e.g., the $\beta$ Pictoris moving
group, TW Hya Association; \citealt{zuckerman2004};
\citealt{torres2008}).  These typically have tens of known members,
with masses $\sim$0.02-2~\msun.  While these are extremely useful due
to their proximity, studies based on these samples may be subject to
small number statistics.  Fortunately, there exist large collections
of young stars, only slightly more distant, in nearby OB associations.
OB associations contain members across the mass spectrum, including
the hot and massive O- and B-type stars for which they are named,
intermediate-mass A/F stars, lower-mass G/K/M stars, and very low-mass
free-floating substellar objects.  Though the lower-mass ($<$ 1.5
\msun) stars blend in with the Galactic field population and are
therefore much more difficult to identify than the OB stars, they
comprise the dominant stellar component of OB associations, typically
present in the thousands \citep{briceno2007}.

Scorpius-Centaurus (Sco-Cen), the nearest OB association, has been the
subject of several efforts to identify its lower-mass population
\citep[e.g.,][]{dezeeuw1999,preibisch1999,mamajek2002,rizzuto2011,rizzuto2015}.
Sco-Cen harbors a barely explored population of thousands of low-mass
K and M-type stars \citep{preibisch2008}.  The association consists of
three classic subregions first defined by \citet{blaauw1946}, and later
refined by \citet{dezeeuw1999}: Upper Scorpius (US), Upper
Centaurus-Lupus (UCL), and Lower Centaurus-Crux (LCC).  The subgroups
have mean distances of 145~pc, 140~pc, and 118~pc, respectively
\citep{dezeeuw1999}.  Though many K/M type members of US have been
identified in surveys by
\cite{walter1994}, \cite{preibisch1998}, \cite{preibisch1999},
\cite{preibisch2001}, \cite{preibisch2002}, and \cite{rizzuto2015},
UCL and LCC occupy larger regions of the sky, and have received less
attention, with only $\sim$90 K/M stars identified in both subgroups,
mostly in \citet{mamajek2002}, \citet{preibisch2008}, 
\citet{song2012}, and a few new M-type members discovered by
\cite{rodriguez2011} and \cite{murphy2015} using Galaxy Evolution 
Explorer (GALEX) UV observations.  A deep DECam imaging survey of UCL 
and LCC is underway which is yielding dozens of members down to the
deuterium-burning limit (Moolekamp et al., in prep).

In this study, we identify and characterize a new sample of K/M-type
($\sim$0.7-1.3~\msun) members of Sco-Cen selected through their X-ray
emission and proper motions.  We describe the results of a
low-resolution spectroscopic survey to identify new low-mass K- and
M-type members of Sco-Cen.  We combine our newly identified members
with members discovered by previous surveys to estimate the accretion
disk fraction, and probe the star-formation history of each subgroup.
Finally, we place these results in context with the results from other
stellar populations.

\section{Sample Selection}

We aim to find new K- and M- type members of all three subgroups of
Sco-Cen.  Low-mass pre-main sequence stars are X-ray luminous, with
log(L$_x$/L$_{bol}$) $\simeq$ -3, due to their strong magnetic dynamos
and coronal X-ray emission \citep{feigelson1999}.  To build our
candidate star list, we cross-referenced the PPMX proper-motion
catalog \citep{roser2008} with X-ray sources within 40\arcsec\, in the
ROSAT Bright Source \citep{voges1999} and Faint Source
\citep{voges2000} catalogs.  We restricted our search to the
\citet{dezeeuw1999} boundaries for Sco-Cen and adopted their proper
motion limits with an extra 10~\masyr\, to allow for larger proper
motion uncertainties and possible kinematic sub-structure.  For
objects at a distance of $\sim$140~pc, this extra 10~\masyr\,
translates to $\simeq 7$~km~s$^{-1}$.  For US, UCL, and LCC this gives
limits of $\mu$ $<$ 47~\masyr, 12~\masyr $<$ $\mu$ $<$ 55~\masyr, and
15~\masyr $<$ $\mu$ $<$ 55~\masyr, respectively, with $\mu_{\alpha} <$
10 \masyr\, and $\mu_{\delta} <$ 30 \masyr\, for all three subgroups.
We also required that our proper motions were less than 50\% uncertain,
$\sigma_{\mu} < 0.5 \mu$, to avoid candidates with poorly constrained 
proper motions.  The mean proper motion magnitude varies across the 
association due to its large extent on the sky and will also vary 
for stars of different distances, this being the principle behind 
kinematic parallaxes.
The intrinsic J-K$_S$ color of an unreddened K0V dwarf is 0.48 mag
\citep{pecaut2013}, and we aimed to find low-mass K- or M-type members
of Sco-Cen so we further made color-magnitude cuts and required our
candidates to have color J-K$_S$ $>$ 0.5 mag and magnitude 
7.0 $<$ J $<$ 11.0 from the 2MASS Point Source Catalog 
\citep{skrutskie2006}.  To avoid very large integration times on the 
SMARTS 1.5m telescope, we chose $J < 11.0$ mag, which corresponds to 
$V \sim 12.5$ mag for an unreddened K0V dwarf.  Stars brighter than 
$J \sim 7.0$ mag would have been covered by previous surveys 
(e.g., \citealt{mamajek2002}), so we chose $J > 7.0$ mag.  At the time 
we performed our sample selection, accurate V-band magnitudes were not 
available for most stars in our parent sample, so we used J-band 
magnitudes to filter the brighter and fainter ends of our parent 
sample.  We removed candidates located in UCL just below US, in 
$343 < l < 350$ and $0 < b < 10$ in galactic coordinates.  This region 
overlaps with ``Lower Sco'' 
\citep{mamajek2013aas,nguyen2013ppvi,mamajek2013ppvi}, and will be 
discussed in forthcoming papers (Mamajek et al., in prep.; 
Nguyen et al., in prep).  This left us with 677 candidates, listed in 
\autoref{tbl:inputdata}.  Before assembling a list of targets for 
observation, we searched the literature to record those stars which had 
been studied in previous surveys from which spectral type, Li or 
H$\alpha$ measurements were available in sufficient detail to make a 
membership determination.  We then omitted stars which were studied in 
\citet{mamajek2002},
\citet{preibisch1998}, \citet{preibisch2001}, \citet{preibisch2002},
\citet{ardila2000}, \citet{koehler2000}, \citet{slesnick2006},
\citet{krautter1997}, \citet{wichmann1997}, \citet{riaz2006}, and
\citet{torres2006}.  This left us with 365 candidates in our
spectroscopic target list.  Objects PPMX~J121431.8-511015,
PPMX~J134751.4-490148, PPMX~J143751.3-545649, and
PPMX~J154348.8-392737 were not observed because they were within
60$\arcsec$ of known Sco-Cen members. The first two may constitute new
companions to Sco-Cen members MML~9 and MML~38, respectively (listed 
in \autoref{tbl:binary}).  The status of the third (PPMX~J143751.3-545649)
with respect to MML~47 is unclear. The fourth, PPMX~J154348.8-392737,
is a previously catalogued companion to HD~140197 \cite[listed as SEE
  247 AB in the Washington Double Star catalog;][]{mason2001}.  This 
finally leaves 361 candidates which were spectroscopically observed.

\onecolumn
\begin{table}
\caption{Photometry and proper motion data for candidates in the Sco-Cen region}\label{tbl:inputdata} 
\begin{tabular}{
@{\hspace{0mm}}l @{\hspace{1.5mm}}c @{\hspace{1.5mm}}c @{\hspace{1.5mm}}c @{\hspace{1.5mm}}c @{\hspace{1.5mm}}c @{\hspace{1.5mm}}c
@{\hspace{1.5mm}}c @{\hspace{1.5mm}}c @{\hspace{1.5mm}}c @{\hspace{1.5mm}}c @{\hspace{1.5mm}}c}
\hline
2MASS &
$\mu_{\alpha}$ & 
$\mu_{\delta}$ & 
Ref. & 
$V$ & 
Ref. & 
$B$--$V$ & 
Ref. & 
$J$ & 
$H$ & 
$K_S$ & 
Note \\
 & 
(\masyr) & 
(\masyr) & 
 &
(mag) & 
 &
(mag) & 
 & 
(mag) & 
(mag) & 
(mag) & 
  \\
\hline
10004365-6522155 & -11.7 $\pm$ 1.9 & 12.4 $\pm$ 1.9 & PX & 10.972 $\pm$ 0.046 & A7 & 0.650 $\pm$ 0.077 & A7 & 9.614 $\pm$ 0.026 & 9.193 $\pm$ 0.026 & 9.018 $\pm$ 0.019 & \\
10065573-6352086 & -19.6 $\pm$ 1.5 & 10.3 $\pm$ 1.8 & U4 & 10.950 $\pm$ 0.012 & A7 & 0.862 $\pm$ 0.018 & A7 & 9.262 $\pm$ 0.028 & 8.744 $\pm$ 0.061 & 8.580 $\pm$ 0.024 & \\
10092184-6736381 & -14.6 $\pm$ 1.8 & 19.6 $\pm$ 1.0 & U4 & 11.512 $\pm$ 0.017 & A7 & 0.849 $\pm$ 0.051 & A7 & 10.027 $\pm$ 0.026 & 9.497 $\pm$ 0.024 & 9.382 $\pm$ 0.021 & \\
10111521-6620282 & -26.5 $\pm$ 1.6 & 11.3 $\pm$ 1.6 & U4 & 12.242 $\pm$ 0.028 & A7 & 0.825 $\pm$ 0.035 & A7 & 10.680 $\pm$ 0.024 & 10.223 $\pm$ 0.026 & 10.073 $\pm$ 0.023 & \\
10293275-6349156 & -11.3 $\pm$ 1.4 & 14.7 $\pm$ 1.5 & U4 & 11.373 $\pm$ 0.014 & A7 & 0.822 $\pm$ 0.027 & A7 & 9.818 $\pm$ 0.022 & 9.349 $\pm$ 0.022 & 9.290 $\pm$ 0.019 & \\
\hline
\end{tabular}
\begin{flushleft}
Adopted 2MASS $JHK_S$ magnitudes are PSF-fit photometry unless otherwise specified;
(a) 2MASS $JHK_S$ aperture photometry; 
(b) 2MASS $JHK_S$ 6x catalog \citep{cutri2012a}; 
(c) 2MASS $H$ psf photometry / $JK_S$ aperture photometry; \\
References -- 
(PX) PPMX, \cite{roser2008}; 
(U4) UCAC4, \cite{zacharias2013}; 
(T2) Tycho-2, \cite{hog2000}; 
(A6) APASS DR6 \citep{henden2012}; 
(A7) APASS DR7 \citep{henden2012};
(HP) Hipparcos, \cite{esa1997}; 
(W97) \cite{wichmann1997}; 
(T06) \cite{torres2006}; \\
Only the first five rows are shown; this table is available in its entirety in the electronic version of the journal.
\end{flushleft}
\end{table}
\twocolumn

\onecolumn
\begin{table}
\caption{New Candidate Double Stars in Sco-Cen}\label{tbl:binary}
\begin{tabular}[cb]{llllllll}
\hline
Primary  & Primary         & Secondary       & J        & PA     & sep.    & epoch     & Notes\\
Name     & PPMX            & PPMX            & mags     & deg    & \arcsec & yr        & \\
\hline
MML 9    & 121434.0-511012 & 121431.8-511015 & 8.7,10.0 & 261.34 & 21.17 & 2010.5589 & 1,2\\
MML 38   & 134750.5-490205 & 134751.4-490148 & 9.3,10.1 & 27.36  & 19.01 & 2010.5589 & 2\\
\hline
\hline
\end{tabular}
\begin{flushleft}
Notes:
(1) The RAVE 4th data release \citep{kordopatis2013} reports radial
velocities of 12.6\,$\pm$\,3.5 km\,s$^{-1}$ and 11.3\,$\pm$\,3.6
km\,s$^{-1}$ for the primary and secondary, respectively, supporting
the notion that they constitute a physical pair.
(2) Astrometry from AllWISE positions \citep{cutri2014}. 
\end{flushleft}
\end{table}
\twocolumn

\section{Observations and Data}

\subsection{Spectra}

Low-resolution red ($\sim$5600\angstrom--6900\angstrom) optical
spectra were obtained from the SMARTS 1.5m telescope at Cerro Tololo
Inter-American Observatory (CTIO), Chile.  Observations were made in
queue mode with the RC spectrograph between July 2009 and October
2010, and in classical mode during the nights of UT 7-17 April 2010.
The spectra were taken with the ``47/Ib'' setup, which consists of a
grating with groove density of 831 grooves~mm$^{-1}$, blaze wavelength
7100\angstrom, a GG495 filter and a slit width of 110.5$\mu$m, giving
a spectral resolution of $\sim$3.1~\angstrom\, in the red optical.
One comparison arc of Ne was taken immediately before three
consecutive exposures of each target.  The data were reduced using the
SMARTS RC Spectrograph IDL pipeline written by Fred
Walter\footnote{\url{http://www.astro.sunysb.edu/fwalter/SMARTS/smarts_15msched.html\#RCpipeline}}.
The three object images are median filtered, bias-trimmed, overscan-
and bias-subtracted, and flat-fielded.  The spectrum is
wavelength-calibrated using the Ne comparison frames.  Finally, we
normalize the spectra to the continuum with a low order spline in
preparation for spectral classification.

In addition, we reanalyzed the low resolution spectra from the \citet{mamajek2002} 
study.  These blue and red spectra were taken at Siding Springs 
Observatory in UT 20-24 April 2000.  The blue spectra have spectral resolution of 
$\sim$2.8\angstrom\, with spectral coverage of $\sim$3850-5400\angstrom.  The red 
spectra have spectral resolution of $\sim$1.3\angstrom\, with spectral coverage of 
$\sim$6200-7150\angstrom.  Further details regarding these spectra are described 
in \cite{mamajek2002}.

\subsection{Photometry}
Our compiled photometry is listed in \autoref{tbl:inputdata}.
Six of our candidates have $BV$ photometry available from the {\it Hipparcos}
catalog, and we adopted it where possible.  For $\sim$130 candidates, we 
adopted $BV$ photometry from the Tycho-2 Catalog \citep{hog2000}, converted 
to the Johnson system using the conversions in \cite{mamajek2002,mamajek2002e}.
For $\sim$350 candidates, we adopted $BV$ photometry from the AAVSO Photometric 
All-Sky Survey\footnote{\url{http://www.aavso.org/apass}} (APASS) Data Release 
6 and Data Release 7 \citep{henden2012}.
and the Search for Associations Containing Young stars catalog 
\cite[SACY;][]{torres2006} for $\sim$100 candidates.   For stars
with V$\simless$12~mag, conservative estimates for SACY photometric uncertainties 
are 0.01~mag (C.A.O. Torres, 2012 private communication).
We adopt JHK$_S$ photometry from the Two Micron All-Sky Survey Point Source 
Catalog \citep[2MASS;][]{skrutskie2006}.  We adopt 2MASS aperture photometry when 
the PSF is poorly fit as indicated by the quality flags (``qflg'' other than `A')
in the 2MASS catalog.  These data are indicated in the notes of 
\autoref{tbl:inputdata}.

For candidate members, we use mid-IR photometry from the Wide-field Infrared 
Survey Explorer \citep[{\it WISE};][]{wright2010,cutri2012b}.  We visually 
examined all images of members which exhibit infrared excesses (identified in 
section~\ref{sec:irexcesses}) in one or more WISE bands.  We flagged photometry 
which could be affected by blends with adjacent objects, unresolved binaries, 
or extended emission from nearby bright stars.  We exclude photometry with 
unreliable detections or uncertainties exceeding 0.25~mag.  We followed the 
scheme for evaluating the photometry in \cite{luhman2012}.  Our mid-IR WISE 
photometry is listed in \autoref{tbl:irdata}.

A few stars in our sample had WISE photometry contaminated by nearby objects.
PDS~415 and CD-23~12840 are a $\sim$4\arcsec\, binary, but only CD-23~12840 is in the 
PPMX catalog and thus only CD-23~12840 is in our sample.  Both components of the pair 
are listed in the WISE All-Sky catalog but CD-23~12840 has $K_S - W1 < 0.0$.  
Examination of the WISE images shows the two are blended. PDS~415 is detected as an 
infrared source in the AKARI catalog \citep{ishihara2010} but CD-23~12840 is not 
detected.  Thus we conclude that PDS~415 is the source of the infrared excess and 
the CD-23~12840 $K_S$ band excess may be spurious.  We exclude the CD-23~12840 WISE 
photometry.  HD~326277 is blended with a nearby object in $W1$ and $W2$ and had 
no reliable detection in $W3$ and $W4$.  HD~326277 has {\it Spitzer} 3.6$\mu$m and 
4.5$\mu$m IRAC photometry in {\it GLIMPSE}
([3.6]=8.165$\pm$0.040~mag, [4.5]=8.178$\pm$0.044~mag; \citealt{churchwell2009}) 
which differs from the WISE $W1$ and $W2$ by $\sim$0.5~mag. 
This suggests that the WISE photometry for this star is contaminated by 
blends and therefore we exclude the WISE photometry for HD~326277.  

\subsection{Astrometry}
Although our original selection scheme utilized proper motions from
the PPMX catalog, we adopt proper motions for our analysis from
several catalogs, including the {\it Hipparcos} catalog
\citep{vanleeuwen2007}, the Tycho-2 catalog \citep{hog2000}, the PPMX
catalog \citep{roser2008} and the UCAC4 catalog \citep{zacharias2013}.
At the time we performed our sample selection, the PPMX catalog 
the most complete homogeneous proper motion catalog with absolute 
proper motions (i.e., on the International Celestial Reference 
System, or ICRS; \citealt{roser2008}).  However, for our analysis we 
make use of proper motions for kinematic distance estimates, and 
therefore desire the most precise and well-constrained proper motions
available.  Therefore, for our calculations, we select the source 
catalog on a case-by-case basis, adopting the proper motions which 
have the smallest uncertainties.  The majority of our adopted proper 
motions listed in \autoref{tbl:inputdata} are from the UCAC4 
catalog.

\section{Analysis}

\subsection{Spectral Classification} \label{sec:MK}
The optical spectra were visually classified against a dense grid of carefully 
chosen spectral 
standards\footnote{see \url{http://www.pas.rochester.edu/~emamajek/spt/}} 
with spectral coverage from $\sim$5600\angstrom--6900\angstrom.  We use 
the same spectral standards and classification criteria as described in 
\cite{pecaut2013}.  The G/K stars are on the classification system of 
\citet{keenan1989} and the M-type stars are on the classification system of 
\citet{kirkpatrick1991}.  

While estimating temperature types for our sample we ignored the Na~I
doublet at $\sim$5889/5896\angstrom\, because it increases in strength
with surface gravity, and is thus useful in discriminating between
dwarfs and giants.  For those pre-main sequence (pre-MS) members of
Sco--Cen, we expected these to have a Na~I doublet line similar to but
weaker than dwarfs \citep{spinrad1962,lawson2009,schlieder2012b}.
Once a temperature type had been established, we compared the Na~I
doublet to that of a dwarf and a giant of the same temperature
subclass, assigning an appropriate luminosity class.  In a few cases
the star had a Na~I doublet feature which closely resembled a giant
(luminosity class III) but the relative strength of the Ca~I at
$\lambda$ 6102, $\lambda$6122 and $\lambda$6162 lines relative to the
Fe~I line at $\lambda$6137 resembled that of a dwarf (luminosity class
V).  In these cases we assigned the intermediate luminosity class of a
subgiant (luminosity class IV).  
For the early to mid G-type stars, the Na~I doublet for subgiants is 
nearly indistinguishable from that of dwarfs (see 
\autoref{fig:teff_na}), so the Na~I doublet was only used to assign
luminosity classes from spectral type $\sim$G5 to $\sim$M3.

\onecolumn
\begin{table}
\caption{Infrared photometry and Infrared Excesses for Members of Sco-Cen} \label{tbl:irdata}
\begin{tabular}{
@{\hspace{0mm}}l @{\hspace{1.0mm}}l @{\hspace{1.0mm}}c @{\hspace{1.0mm}}c @{\hspace{1.0mm}}c @{\hspace{1.0mm}}c 
@{\hspace{1.0mm}}c @{\hspace{1.0mm}}c @{\hspace{1.0mm}}c @{\hspace{1.0mm}}c @{\hspace{1.0mm}}c }
\hline
2MASS &
WISE &
$W1$ &
$W2$ &
$W3$ &
$W4$ &
$E(K_S-W1)$ &
$E(K_S-W2)$ &
$E(K_S-W3)$ &
$E(K_S-W4)$ \\
J &
J &
(mag) &
(mag) &
(mag) &
(mag) &
(mag) &
(mag) &
(mag) &
(mag) \\
\hline
10065573-6352086 & 100655.70-635208.6 & 8.513 $\pm$ 0.023 & 8.531 $\pm$ 0.021 & 8.446 $\pm$ 0.018 & 8.301 $\pm$ 0.131 & -0.025 $\pm$ 0.033 & -0.006 $\pm$ 0.032 & 0.041 $\pm$ 0.030 & 0.115 $\pm$ 0.133 \\ 
10092184-6736381 & 100921.83-673637.9 & 9.056 $\pm$ 0.019 & 9.177 $\pm$ 0.018 & 9.201 $\pm$ 0.020 & 8.916 $\pm$ 0.190 & 0.233 $\pm$ 0.028 & 0.142 $\pm$ 0.028 & 0.078 $\pm$ 0.029 & 0.291 $\pm$ 0.191 \\ 
10313710-6901587 & 103137.09-690158.7 & 9.414 $\pm$ 0.023 & 9.431 $\pm$ 0.021 & 9.360 $\pm$ 0.025 & 9.622 $\pm$ 0.435 & -0.013 $\pm$ 0.034 & -0.011 $\pm$ 0.033 & 0.013 $\pm$ 0.035 & -0.325 $\pm$ 0.436 \\ 
10334180-6413457 & 103341.79-641345.6 & 9.236 $\pm$ 0.025 & 9.218 $\pm$ 0.021 & 9.167 $\pm$ 0.030 & ... & -0.038 $\pm$ 0.035 & -0.001 $\pm$ 0.032 & 0.003 $\pm$ 0.038 & ... \\ 
10412300-6940431 & 104122.96-694042.8 & 8.304 $\pm$ 0.022 & 8.330 $\pm$ 0.020 & 8.216 $\pm$ 0.015 & 7.858 $\pm$ 0.082 & -0.008 $\pm$ 0.030 & 0.012 $\pm$ 0.028 & 0.093 $\pm$ 0.025 & 0.382 $\pm$ 0.084 \\ 
\hline
\end{tabular}
\begin{flushleft}
All infrared photometry ($W1$, $W2$, $W3$ and $W4$) above is adopted from the {\it WISE} catalog \citep{cutri2012b}.
Only the first five rows are shown; this table is available in its entirety in the electronic version of the journal.
\end{flushleft}
\end{table}
\twocolumn

We also revised the spectral classifications for Sco-Cen members
studied in \citet{mamajek2002}.  
The spectral types from \citet{mamajek2002} were tied very closely to
those of the Michigan Spectral Survey \citep{houk1978, houk1982}.  The
\citet{mamajek2002} spectra were re-classified by M. Pecaut during the
survey of \citet{pecaut2013} using the standards of \citet{keenan1989},
and systematic differences were noted.  In \autoref{tbl:membership} 
we adopt the revised classifications for the \citet{mamajek2002} stars.  
There are clearly systematic differences between the Michigan types 
and those on the modern MK system \cite[as noted in Appendix C.1 of
][]{pecaut2013}. Put simply, a G2V star classified in the Michigan
survey corresponds more closely to a G0.5V on the modern grid of
G-dwarf standards \citep{keenan1989}, as classified by Gray and
collaborators \citep[e.g.][]{gray2003, gray2006}. This is verified by
comparison of the colors and effective temperatures of stars
classified both by Houk and Gray et al. While the Michigan survey
mostly relied on the MK system of \citet{johnson1953} (including changes
up through early 1970s), there were minor shifts to the MK system for
stars hotter than G0 by Morgan \citep{morgan1973a,morgan1978}, and for
GK-type stars by \citet{keenan1989} after the Michigan survey was
initiated in the late 1960's. The modern M dwarf classifications rely
on the standard sequence of \citet{kirkpatrick1991} and its later
additions \citep[e.g.][]{henry2002}. The spectral classification surveys
of bright stars undertaken by Gray and collaborators \citep{gray2001a,
gray2003, gray2006} are based on the last generations of Morgan's and
Keenan's hot and cool star spectral sequences (\citealt{morgan1978},
\citealt{keenan1989}) which are in common modern use.  We compared stars 
of given spectral type in the Michigan survey to those of 
\citet{gray2001a, gray2003, gray2006}, and in \autoref{tbl:MK} we 
provide an estimate of the modern dwarf spectral type for given Michigan 
spectral types.  Very few of the Michigan spectral classes appear to 
correspond closely to the same type of star on the modern MK grid (e.g. 
F6V, K0V, K3V being rare exceptions). Unfortunately, the differences are 
most pronounced in the early G-type dwarfs, where Michigan G2V 
corresponds to Gray's G0.5V, Michigan G3V corresponds to G1.5V, etc.  
This is partially due to Houk's choice of G dwarf standards (e.g. 
using $\beta$ Com as their G2V standard, despite it being considered a 
G0V or F9.5V standard elsewhere), but also due to Keenan's minor
adjustments to the G dwarf standards throughout the 1980's. As the
\citet{keenan1989} grid defines the modern GK dwarf sequence and is
common use, classifications that used this grid are to be considered
on the modern MK system following \autoref{tbl:MK}.  For 
intercomparison of samples of stars classified in the Michigan survey 
with samples classified on the modern MK system, we recommend 
converting the Michigan spectral types to the modern MK system using 
the conversions in \autoref{tbl:MK}.  None map exactly to G2V.

\onecolumn
\begin{table}
\caption{Membership properties for Sco-Cen candidates} \label{tbl:membership}
\begin{tabular}{lrlrrllllrll}
\hline
2MASS &
SpT &
Ref. &
$\mathrm{EW(Li)}$ & 
Ref. &
\multicolumn{5}{c}{Membership} & 
Young Star &
Sco-Cen \\
\cline{6-10} 
 & 
 & 
 & 
(\angstrom) & 
 &
Li? & 
$\mu$? & 
$\log(g)$? & 
HRD? & 
Final &
Ref. &
Ref. \\
\hline
10004365-6522155 & G3V & PM & 0.24 & PM & Y & N & Y? & ... & N &  ...  &  ...  \\
10065573-6352086 & K0Ve & T06 & 0.35 & T06 & Y & Y & ... & Y & Y & T06 & PM \\
10092184-6736381 & K1IV(e) & PM & 0.32 & PM & Y & Y & Y? & Y & Y & PM & PM \\
10111521-6620282 & K0IV & PM & 0.33 & PM & Y & Y & Y? & N & N &  ...  &  ...  \\
10293275-6349156 & K1IV & PM & 0.23 & PM & Y? & N & Y? & ... & N &  ...  &  ...  \\
\hline
\end{tabular}
\begin{flushleft}
References in the final two columns refer to the first work to identify the star
as young (``Young Star Ref.'') and the first work to identify it as a member of
Sco-Cen (``Sco-Cen Ref.''). \\
References --
(PM) - This work;
(S49) - \citet{struve1949};
(H54) - \citet{henize1954};
(D59) - \citet{dolidze1959};
(T62) - \citet{the1962};
(T64) - \citet{the1964};
(G67) - \citet{garrison1967};
(S72) - \citet{satyvaldiev1972};
(F75) - \citet{filin1975};
(P75) - \citet{petrov1975};
(H76) - \citet{henize1976};
(S77) - \citet{schwartz1977};
(S82) - \citet{satyvoldiev1982};
(S86) - \citet{stephenson1986b};
(F87) - \citet{finkenzeller1987};
(D88) - \citet{downes1988};
(He88) - \citet{herbig1988};
(Ho88) - \citet{houk1988};
(I89) - \citet{ichikawa1989};
(M89) - \citet{mathieu1989};
(S91) - \citet{stocke1991};
(V91) - \citet{vazquez1991};
(B92) - \citet{bouvier1992};
(C92) - \citet{carballo1992};
(G92) - \citet{gregorio-hetem1992};
(H94) - \citet{hughes1994};
(W94) - \citet{walter1994};
(R95) - \citet{randich1995};
(B96) - \citet{brandner1996};
(P96) - \citet{park1996};
(F97) - \citet{feigelson1997};
(K97) - \citet{krautter1997};
(W97) - \citet{wichmann1997};
(M98) - \citet{martin1998};
(P98) - \citet{preibisch1998};
(S98) - \citet{sciortino1998};
(D99) - \citet{dezeeuw1999};
(P99) - \citet{preibisch1999};
(W99) - \citet{wichmann1999};
(H00) - \citet{hoogerwerf2000};
(K00) - \citet{koehler2000};
(W00) - \citet{webb2000};
(P01) - \citet{preibisch2001};
(Z01) - \citet{zuckerman2001b};
(M02) - \citet{mamajek2002};
(P02) - \citet{preibisch2002};
(R03) - \citet{reid2003};
(M05) - \citet{mamajek2005};
(T06) - \citet{torres2006};
(R06) - \citet{riaz2006};
(Ri06) - \citet{riaud2006};
(C06) - \citet{carpenter2006};
(W07) - \citet{white2007};
(V09) - \citet{vianaalmeida2009};
(K11) - \citet{kiss2011};
(R11) - \citet{rodriguez2011};
(L12) - \citet{luhman2012};
(M12) - \citet{mamajek2012a};
(S12) - \citet{song2012};
(R15) - \citet{rizzuto2015};
(A)   - measured from spectrum downloaded from the ESO archive, program 083.A-9003(A);
(B)   - measured from spectrum downloaded from the ESO archive, program 081.C-0779(A)	 \\
Only the first five rows are shown; this table is available in its entirety in the electronic version of the journal.
\end{flushleft}
\end{table}
\twocolumn

\begin{table}
\caption{Comparing Michigan Spectral Types to Gray MK Spectral Types}\label{tbl:MK}
\setlength{\tabcolsep}{0.03in}
\begin{tabular}{lll|lll|lll}
\hline
{(1)} & & {(2)}  & {(3)} & &{(4)} &{(5)} & &{(6)} \\
{SpT} & & {SpT}  & {SpT} & &{SpT} &{SpT} & &{SpT} \\
{Houk}& & {Gray} & {Houk}& &{Gray}&{Houk}& &{Gray} \\
\hline
F0V    & $\rightarrow$ & F1V   & G0V   & $\rightarrow$ & F9.5V    & K0V    & $\rightarrow$ & K0V \\
F1V    & $\rightarrow$ & F2V   & G0/1V & $\rightarrow$ & F9.5V    & K0/1V  & $\rightarrow$ & K1V \\
F2V    & $\rightarrow$ & F3.5V & G0/2V & $\rightarrow$ & G3:V     & K1V    & $\rightarrow$ & K1.5V \\
F2/3V  & $\rightarrow$ & F3.5V & G1V   & $\rightarrow$ & G0+V     & K1/2V  & $\rightarrow$ & K2V \\
F3V    & $\rightarrow$ & F4V   & G1/2V & $\rightarrow$ & G0V      & K2V    & $\rightarrow$ & K2.5V \\
F3/5V  & $\rightarrow$ & F5V   & G2V   & $\rightarrow$ & G0.5V    & K2/3V  & $\rightarrow$ & K3V \\
F5V    & $\rightarrow$ & F5.5V & G2/3V & $\rightarrow$ & G1V      & K3V    & $\rightarrow$ & K3V \\
F5/6V  & $\rightarrow$ & F6V   & G3V   & $\rightarrow$ & G1.5V    & K3/4V  & $\rightarrow$ & K4V \\
F6V    & $\rightarrow$ & F6V   & G3/5V & $\rightarrow$ & G4V      & K4V    & $\rightarrow$ & K4+V \\
F6/7V  & $\rightarrow$ & F6.5V & G5V   & $\rightarrow$ & G4.5V    & K4/5V  & $\rightarrow$ & K5V \\
F7V    & $\rightarrow$ & F7.5V & G5/6V & $\rightarrow$ & G6V      & K5V    & $\rightarrow$ & K6-V \\
F7/8V  & $\rightarrow$ & F8V   & G8V   & $\rightarrow$ & G8.5V    & K5/M0V & $\rightarrow$ & K7V \\
F8V    & $\rightarrow$ & F8.5V &       &               &          & M0V    & $\rightarrow$ & K6.5V \\
F8/G0V & $\rightarrow$ & F9V   &       &               &          &        &               &     \\
\hline
\hline
\end{tabular}
\end{table}
\twocolumn

\subsection{Distance Calculation\label{sct:dist}}

Very few of our candidate members have measured trignometric
parallaxes, so we estimate distances to each candidate by calculating
a ``kinematic'' or ``cluster'' parallax
\citep[e.g.,][]{debruijne1999a}.  This method uses the centroid space
motion of the group with the proper motion of the candidate member and
the angular separation to the convergent point of the group to
estimate the distance to the individual member, with the assumption
that they are co-moving.  We emphasize that kinematic parallaxes are
only meaningful for true members and meaningless for non-members.
Kinematic parallaxes have been shown to reduce the scatter in the H-R
diagram for cluster members over trignometric parallaxes
\citep{debruijne1999b}.  We adopt the formalism and methods of
\cite{mamajek2005} and adopt the updated Sco-Cen subgroup space
motions listed in \citet{chen2011}.  In addition to providing improved
distance estimates over simply adopting the mean subgroup distances,
the kinematic parallaxes allow us identify non-members and assess the
likelihood that the candidate member is a bona-fide member.

\subsection{Membership Criteria}\label{sec:membership}

In order to identify likely members from our sample, we demand that
all available data paint a consistent picture of association
membership, and therefore consider several indicators to discriminate
against interlopers.  Based on previous surveys \citep[see][and
  references therein]{preibisch2008}, we expect the K and M-type stars
to be pre-MS, and therefore have Li absorption stronger, on average,
than that of a $\sim$30-50~Myr-old population.  We also expect that
they will exhibit surface gravities intermediate between dwarfs and
giants.  We identify those stars which have the following:

\begin{enumerate}

\item Appropriate levels of Li absorption in their spectra, using the
  Li~6708\angstrom\, line,

\item Dwarf or subgiant surface gravities, using the Na~I doublet at
  5889/5896\angstrom,

\item Kinematic distances consistent with Sco--Cen,

\item H-R diagram positions broadly consistent with membership in
  Sco-Cen (e.g., neither below the main sequence nor above a 1~Myr
  isochrone).

\end{enumerate}

We discuss these criteria in detail in the following sections.
Borderline cases were examined closely.

\subsubsection{Lithium}

Pre-MS stars transition from fully convective to having deep
convective envelopes with radiative interiors during their
contraction. These stars can mix Li from the stellar photosphere
throughout the convection zone down to interior regions of the star
where it is destroyed at $T\simgreat$2.6 MK \citep{bodeheimer1965,
strom1994}.  Thus, for $\sim$0.7-1.3~\msun\, stars like those 
targeted in this study, a high photospheric Li abundance is only 
present when the star is very young.  Based on the published ages 
\citep{degeus1989,preibisch1999,mamajek2002,pecaut2012}, we 
expect members of Sco-Cen to exhibit stronger Li absorption lines 
at a given \teff, on average and within some acceptable scatter, 
than a $\sim$30-50~Myr sample.

We measured the equivalent width of the 6708\angstrom\, Li feature for
all stars in our spectroscopic sample.  The values were measured with 
IRAF\footnote{IRAF is distributed by the National
  Optical Astronomy Observatory, which is operated by the Association
  of Universities for Research in Astronomy (AURA) under cooperative
  agreement with the National Science Foundation.}  
with Vogt profiles after the spectrum was normalized to the continuum.
Given our spectral resolution and the repeatability of our 
measurements, we estimate that our reported EW(Li) values are accurate 
to $\sim$0.05\angstrom.  The 6708\angstrom\, feature is a blend with a 
nearby Fe~I line at 6707.44\angstrom, unresolved at our resolution, 
and thus our EW(Li) values are probably overestimated by 
$\sim$0.02\angstrom.  However, this is smaller than our uncertainties
and we do not attempt to correct for this blend.  Many stars in our 
parent sample were not observed because they had EW(Li) values 
available in the literature; for these stars we simply adopted the 
previously published values.

\begin{figure}
\begin{center}
\includegraphics[scale=0.45]{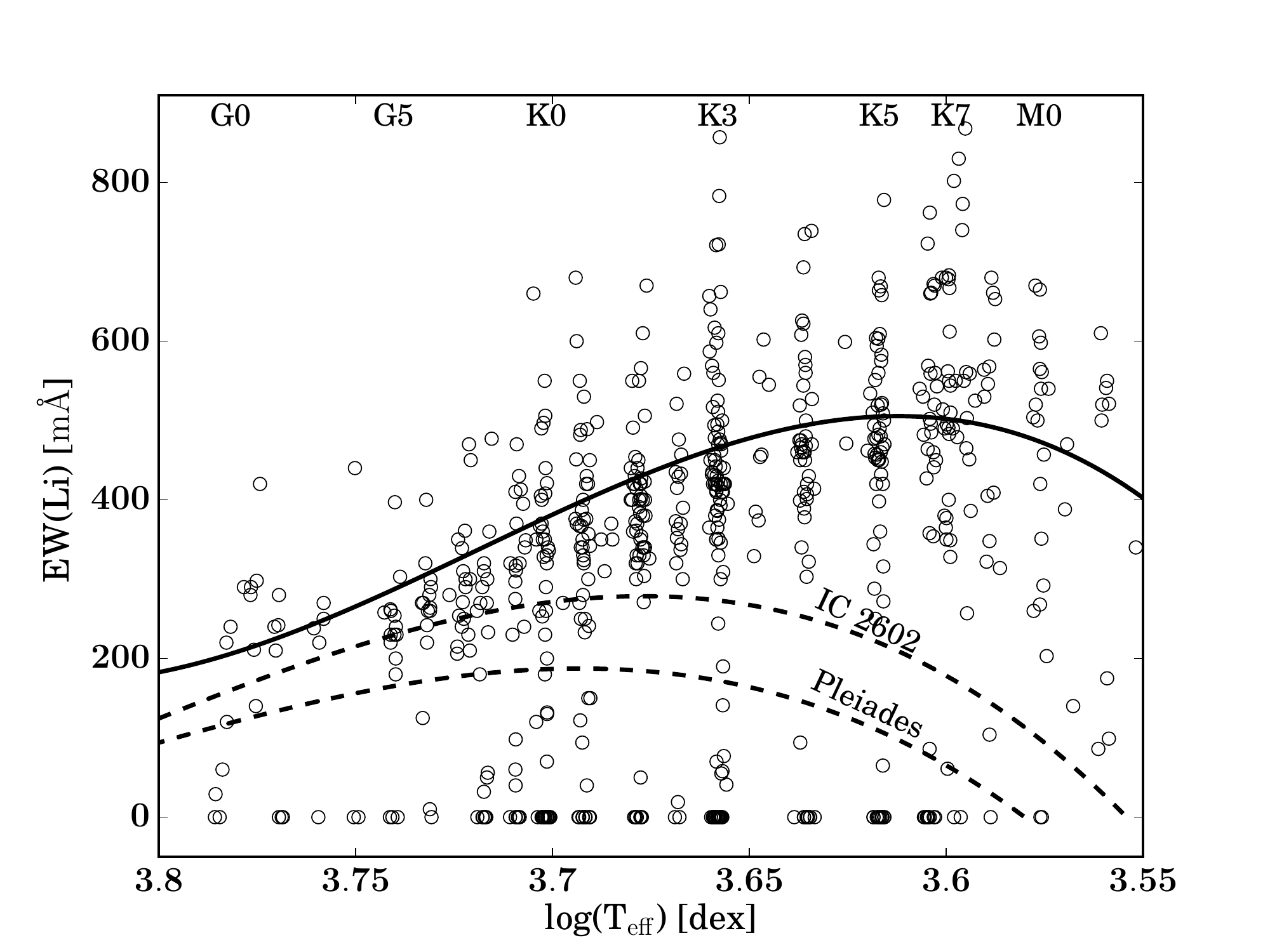}
\caption{
  Measured EW(Li) from the 6708\angstrom\, line plotted against \logt\, 
  for our X-ray sample.
  Plotted dashed lines are polynomial fits to data from surveys in IC~2602 
  \citep{randich1997} and the Pleiades \protect\citep{soderblom1993,jones1996}.  
  The location of the EW(Li) for our Sco-Cen candidate members is used to 
  determine the membership indicator in \autoref{tbl:membership}.  Those above 
  the IC~2602 curve are marked ``Y'', those above the Pleiades curve are marked 
  ``Y?'', while candidates below the Pleiades are marked ``N''.  The solid curve 
  is a polynomial fit to the Sco-Cen candidates marked ``Y''; the polynomial 
  coefficients are listed in \autoref{tbl:lipoly}.
}
\label{fig:teff_li}
\end{center}
\end{figure}

We compare our EW(Li) measurements at a given \teff\, to those in
nearby open clusters.  In \autoref{fig:teff_li} we plot \teff\, 
versus EW(Li) for our sample along with data for the young open 
clusters IC~2602 \citep{randich1997} and the Pleiades
\citep{soderblom1993,jones1996}.  Most stars in our sample exhibit
much larger EW(Li) at a given \teff\, than either IC~2602
($\sim$45~Myr; \citealt{dobbie2010}) or the Pleiades ($\sim$125~Myr;
\citealt{stauffer1998}).  These data suggest that Li is largely
undepleted for the $\sim$10-16 Myr-old stars in Sco-Cen hotter than
spectral type $\sim$K3, but the Li depletion becomes very strong
for stars cooler than $\sim$K3 ($\sim$1.1~\msun).  We fit low-order 
polynomials to the 
IC~2602 and Pleiades \teff\, versus EW(Li) data; the polynomial 
coefficents are listed in \autoref{tbl:lipoly}.  For our Li-based 
membership criterion, if the EW(Li) was above the polynomial fit to 
IC~2602 we marked that star as `Y' in \autoref{tbl:membership}; if 
EW(Li) was below the IC~2602 polynomial fit but above the Pleiades 
polynomial fit, we marked it as `Y?', and if the EW(Li) is below the 
Pleiades polynomial fit we marked it as `N'.  Stars studied in
\citet{preibisch1999}, \citet{koehler2000} and \citet{krautter1997}
were confirmed as Li-rich but the measurements were not reported.  We
marked these candidates as `Y' in \autoref{tbl:membership}.  In total,
we marked 482 candidates as `Y', 153 as `N', and 38 as `Y?'.

\begin{table}
\caption{Polynomial fits to Li vs \teff\, data}\label{tbl:lipoly}
\begin{tabular}{lrrrr}
\hline
Polynomial  & & & \\
Coefficient & IC 2602             & Pleiades                  & Sco-Cen \\
\hline
$a_0$ & -1.869522 $\times$ 10$^6$ & -1.639425 $\times$ 10$^6$ & -3.576491 $\times$ 10$^6$ \\ 
$a_1$ &  1.472855 $\times$ 10$^6$ &  1.290057 $\times$ 10$^6$ &  2.890342 $\times$ 10$^6$ \\ 
$a_2$ & -3.862308 $\times$ 10$^5$ & -3.379855 $\times$ 10$^5$ & -7.778982 $\times$ 10$^5$ \\ 
$a_3$ &  3.371432 $\times$ 10$^4$ &  2.948343 $\times$ 10$^4$ &  6.972998 $\times$ 10$^4$ \\ 
\hline
\hline
\end{tabular}
\begin{flushleft}
Note: $EW(Li) [m\angstrom] = a_0 + a_1 ($\logt$) + a_2 ($\logt$)^2 + a_3 ($\logt$)^3$ \\
Polynomial fit is valid from \logt=3.800~dex to 3.580~dex.
\end{flushleft}
\end{table}

\subsubsection{Surface Gravity}

\begin{figure}
\begin{center}
\includegraphics[scale=0.45]{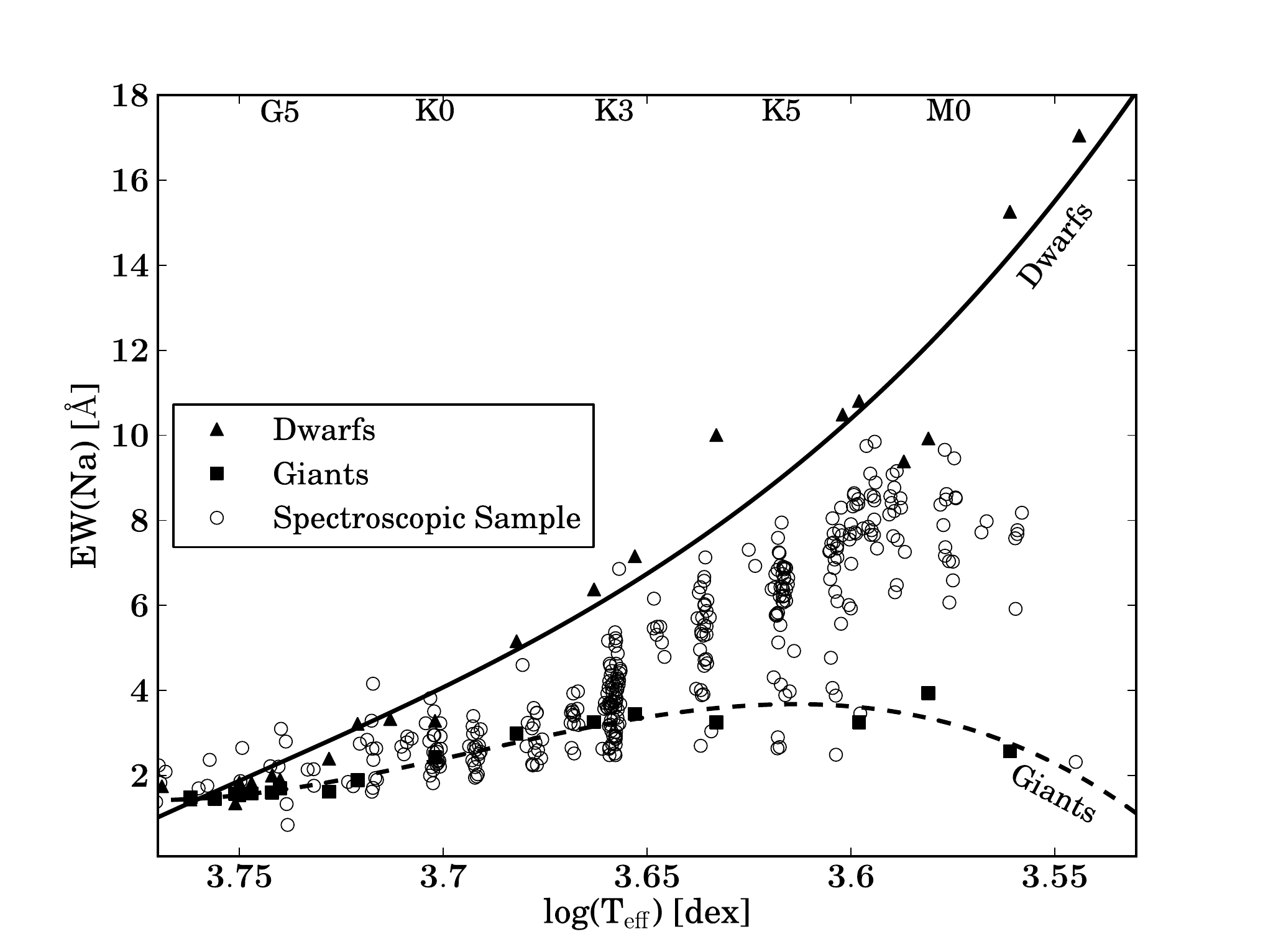}
\caption{Measured EW(Na) from the Na~I doublet at 
  5889\angstrom/5896\angstrom\, line plotted against \logt\, for 
  our spectroscopic sample.  Plotted are polynomial fits to dwarf 
  and giant spectral standard stars used for classification.}
\label{fig:teff_na}
\end{center}
\end{figure}

Based on previous nuclear and pre-MS age determinations of the
subgroups of Sco--Cen \citep{mamajek2002,pecaut2012}, the low-mass
stars in Sco-Cen are expected to be pre-main sequence with surface
gravities intermediate between dwarfs and giants.  For candidates in
our spectroscopic sample we can examine spectral features sensitive to
surface gravity, primarily the Na~I doublet at
5889\angstrom/5896\angstrom.  Details on using Na~I as a surface
gravity indicator for identifying young stars is discussed in detail
in \cite{schlieder2012b}.  The Na~I doublet decreases in strength at a
given \teff\, as surface gravity decreases.  However, the observed
strength of the Na~I doublet in subgiants among G-type stars is
difficult to distinguish from dwarfs.  The differences in the Na~I
doublet strength between dwarfs and subgiants becomes more useful in
mid-K-type spectra, as shown in \autoref{fig:teff_na}.  For this
reason we are fairly conservative in our surface gravity membership
criterion, marking G-type stars with luminosity classes of `IV' with
`Y', `V' with `Y?' and `III' with `N'.  Note that we only use this
criterion for stars which we have observed spectroscopically;
luminosity classifications from other authors for stars in our parent
sample are not used for this criterion.  We assigned 345 stars 
into these two categories, with 300 marked as `Y?' and 45 marked 
as `N'.

\subsubsection{Distances}

As mentioned previously, the majority of our sample lack trigonometric
distance estimates.  However, if we calculate a kinematic parallax
estimate using the best available proper motion, we can then estimate
distances for each star in our sample and verify that this lies within
the range of expected distances for the association.  Stars with
discrepant kinematic distances are identified as non-members.  We use 
the results of the membership study of \citet{rizzuto2011}, which 
makes use of the revised {\it Hipparcos} trignometric parallaxes
\citep{vanleeuwen2007}, to establish our distance criteria.  Based on
the spatial distribution of their candidate members,
\citet{rizzuto2011} concluded that strict boundaries between the
subgroups are somewhat arbitrary, so we do not use distinct kinematic
distance criteria for each subgroup.  To establish parallax ($\varpi$)
limits based on galactic longitude and latitude $(l, b)$, we fit a
plane to the B-type members with membership probabilities $>$50\% from
the \cite{rizzuto2011} study, obtaining:

\begin{eqnarray*}
  \varpi & = & (-0.0208 \pm 0.0061) \times (l - <l>)\\
      &   &  + (-0.0299 \pm 0.0158) \times (b - <b>)  \\
      &   &  + (7.7761  \pm 0.0969)  
\end{eqnarray*}

with mean galactic coordinates $<l>=328.579^{\circ}$, $<b>=13.261^{\circ}$.  
The 1$\sigma$ dispersion in the fit is $\sigma_{\varpi} \simeq 1.25$\,mas.
Therefore we model the shape of Sco-Cen with a continuous plane and
depth characterized by 1$\sigma$ dispersion of $\sim$1.25~mas.  For
our kinematic distance criteria, we mark stars with `Y' in
\autoref{tbl:membership} if their kinematic parallaxes are within
the 2$\sigma$ (95\% C.L.) dispersion from the plane described above,
and `Y?' if they are between 2$\sigma$ and 3$\sigma$.  We mark stars
with `N' if they are beyond 3$\sigma$ of the dispersion from the
plane.  This criteria effectively provides different distance limits
as a function of galactic longitude and latitude, giving typical
2$\sigma$ distance limits of 141$^{+77}_{-37}$~pc,
130$^{+62}_{-32}$~pc, and 117$^{+49}_{-27}$~pc for US, UCL and LCC,
respectively.  677 stars in our sample have been placed in these
categories, with 495 marked as `Y', 80 as `Y?', and 102 as `N'.

\subsubsection{HRD position}

As a final membership check, we place these stars on the H-R diagram
using their kinematic distances, which, for this purpose, {\it
  assumes} they are co-moving with other members in the association.
Objects that have H-R diagram positions inconsistent with membership,
e.g., below the main sequence, are rejected.  Below we discuss the few
stars which were Li-rich and fell inside our kinematic distance
criteria but were rejected because they had discrepant H-R diagram
positions.  Two stars were rejected since they fell below the main 
sequence and two stars were rejected since they were well above the
1~Myr isochrone.  Unresolved binarity is insufficient to account for 
these discrepant positions.  Further analysis can be found in 
section~\ref{sec:hrd_km}.

\subsubsection{Interlopers}

{\it US:} HD~144610 is a Li-rich K1IV-III which has a kinematic
parallax of $\pi_{\rm kin}$=3.41$\pm$0.69~mas which is consistent with
a H-R diagram position of \logt=3.70$\pm$0.01, \logl=1.42$\pm$0.18,
well above the 1~Myr isochrone, so we reject it.

{\it UCL}: 2MASS~J14301461-4520277 is a Li-rich K3IV(e) which, using
the kinematic parallax of $\pi_{\rm kin}=10.92\pm0.92$~mas, has an H-R
diagram position well below the main sequence (\logt=3.66$\pm$0.02,
\logl=-0.91$\pm$0.08), so we reject its membership to UCL.
2MASS~J16100321-5026121 is a Li-rich K0III \citep{torres2006} which
lies well above the 1~Myr isochrone (\logt=3.70$\pm$0.01,
\logl=1.16$\pm$0.08), calculated using the kinematic parallax of
$\pi_{\rm kin}=4.19\pm0.38$~mas, so we consider a UCL interloper.

{\it LCC:} 2MASS~J10111521-6620282, lies well below the main sequence,
with \logt=3.70$\pm$0.01, \logl=-0.86$\pm$0.09, calculated using the
predicted kinematic parallax of $\pi_{\rm kin}$=9.63$\pm$0.94~mas.
This is puzzling because it is a Li-rich K0IV, with a proper motion in
excellent agreement with membership in LCC.  It does not exhibit an
infrared excess which could be a signature of an edge-on disk, and the 
$B-V$ and $V-K_S$ colors are consistent with a negligibly reddened 
young K0.  However, it is $\sim$0.4~dex underluminous for the main 
sequence, so we consider it a likely LCC interloper.

\subsubsection{Final Membership Assessment}

Though we require the Li, surface gravity indicators, kinematic
distance criterion, and H-R diagram criterion to indicate membership,
the Li and kinematic distance criteria are the most restrictive and
are responsible for identifying most non-members.  Stars with `Y' or
`Y?' in each of the four membership categories have been identified 
members of Sco-Cen. Though we have made every effort to remove 
interlopers, our list of candidate members may still contain a few 
non-members, though it will be dominated by true members.  From our 
current sample, we identify 493 stars as likely members, listed in 
\autoref{tbl:member_properties}.  The 180 stars rejected as Sco-Cen 
members are listed in \autoref{tbl:rejected}.  156 are newly 
identified young stars.

\onecolumn
\begin{table}
\caption{Stellar properties for Sco-Cen members} \label{tbl:member_properties}
\begin{tabular}{
@{\hspace{1.0mm}}l @{\hspace{1.0mm}}c @{\hspace{1.5mm}}l @{\hspace{1.5mm}}l @{\hspace{1.5mm}}l
@{\hspace{1.5mm}}r @{\hspace{1.5mm}}l @{\hspace{1.5mm}}r @{\hspace{1.5mm}}r @{\hspace{1.5mm}}r
@{\hspace{1.5mm}}r @{\hspace{1.5mm}}r @{\hspace{1.5mm}}r @{\hspace{1.5mm}}l @{\hspace{1.5mm}}l 
}
\hline
ID & 
2MASS &
Grp. &
SpT &
Ref. &
EW(H$\alpha$) & 
Ref. &
A$_V$ & 
$\pi_{kin}$ & 
\logt & 
\logl & 
Age & 
Mass & 
Disk &
Other Name \\
 & 
 & 
 & 
 & 
 & 
(\angstrom) & 
 & 
(mag) & 
(mas) & 
(dex) & 
(dex) & 
(Myr) & 
(M$_{\odot}$) & 
Type &
\\
\hline
1 & 10065573-6352086 & LCC & K0Ve & T06 & 0.0 & T06 & 0.16$\pm$0.04 & 7.54$\pm$0.83 & 3.702$\pm$0.009 & -0.059$\pm$0.097 & 14 & 1.1 &     & TYC~8951-289-1 \\ 
2 & 10092184-6736381 & LCC & K1IV(e) & PM & -0.2 & PM & 0.00$\pm$0.10 & 7.03$\pm$0.76 & 3.692$\pm$0.012 & -0.334$\pm$0.096 & 29 & 0.9 &     & TYC~9210-1484-1 \\ 
3 & 10313710-6901587 & LCC & K2.5IV(e) & PM & -0.1 & PM & 0.00$\pm$0.13 & 6.91$\pm$0.72 & 3.668$\pm$0.019 & -0.380$\pm$0.095 & 21 & 0.9 &     &  \\ 
4 & 10334180-6413457 & LCC & K2.5IV(e) & PM & 0.0 & PM & 0.00$\pm$0.15 & 5.29$\pm$0.85 & 3.668$\pm$0.019 & -0.077$\pm$0.143 & 7 & 1.2 &     & V542~Car \\ 
5 & 10412300-6940431 & LCC & G8Ve & T06 & 0.0 & T06 & 0.00$\pm$0.02 & 11.05$\pm$0.97 & 3.717$\pm$0.007 & -0.250$\pm$0.078 & 33 & 0.9 &     & TYC~9215-1181-1 \\ 
\hline
\end{tabular}
\begin{flushleft}
EW(H$\alpha$)$<$ 0 denotes emission, EW(H$\alpha$)$>$ 0 denotes absorption. 
We report the Houk type for the star 16102888-2213477 = HD~145208 (G5V), but
internally use the \teff\, and BC for the equivalent Gray type (G4.5V) in our
calculations.  See \autoref{tbl:MK} for more information. \\
References --
(PM) - This work;
(G67) - \citet{garrison1967};
(F87) - \citet{finkenzeller1987};
(Ho88) - \citet{houk1988};
(W94) - \citet{walter1994};
(H94) - \citet{hughes1994};
(W97) - \citet{wichmann1997};
(K97) - \citet{krautter1997};
(P98) - \citet{preibisch1998};
(P99) - \citet{preibisch1999};
(W99) - \citet{wichmann1999};
(K00) - \citet{koehler2000};
(P01) - \citet{preibisch2001};
(M02) - \citet{mamajek2002};
(P02) - \citet{preibisch2002};
(Z01) - \citet{zuckerman2001b};
(T06) - \citet{torres2006};
(R06) - \citet{riaz2006};
(C06) - \citet{carpenter2006};
(W07) - \citet{white2007};
(M11) - \citet{muller2011};
(D12) - \citet{dahm2012};
(A)   - measured from spectrum downloaded from the ESO archive, program 083.A-9003(A);
(B)   - measured from spectrum downloaded from the ESO archive, program 081.C-0779(A); \\
Only the first five rows are shown; this table is available in its entirety in the electronic version of the journal.
\end{flushleft}
\end{table}
\twocolumn

\onecolumn
\begin{table}
\caption{Objects rejected as Sco-Cen members} \label{tbl:rejected}
\begin{tabular}{ @{\hspace{1.5mm}}l @{\hspace{1.5mm}}l @{\hspace{1.5mm}}l @{\hspace{1.5mm}}l @{\hspace{1.5mm}}l }
\hline
2MASS &
SpT &
Ref. &
Rejection &
Other Names \\
 &
 &
 &
Reason &
 \\
\hline
10004365-6522155 & G3V & PM & $\mu$ &      \\ 
10111521-6620282 & K0IV & PM & HRD position &      \\ 
10293275-6349156 & K1IV & PM & $\mu$ & TYC~8964-165-1 \\ 
10342989-6235572 & G7Ve & PM & $\mu$ & HD~307772 \\ 
10441393-6446273 & K1V(e) & PM & $\mu$ & HD~307960 \\ 
\hline
\end{tabular} 
\begin{flushleft}
References--
(PM) - This work;
(K97) - \citet{krautter1997};
(P98) - \citet{preibisch1998};
(K00) - \citet{koehler2000};
(M02) - \citet{mamajek2002};
(T06) - \citet{torres2006}; \\
Only the first five rows are shown; this table is available in its entirety in the electronic version of the journal.
\end{flushleft}
\end{table}
\twocolumn

\subsection{Extinction}

We estimate the reddening and extinction for Sco-Cen members using the
spectral type-intrinsic color sequence for pre-MS stars described in
\citet{pecaut2013}.  These intrinsic colors are calibrated using
low-mass members of nearby moving groups which have negliglible
extinction, and bracket the age range of the Sco-Cen subgroups
(5-30~Myr).  Many of the stars in our sample have very low reddening,
and where we obtained a non-physical negative extinction we set the
extinction to zero.  We used a total-to-selective extinction of
R$_V$=3.1 and calculated E(B-V), E(V-J), E(V-H), and E(V-K$_S$) with
A$_J$/A$_V$=0.27, A$_H$/A$_V$=0.17, A$_{K_S}$/A$_V$=0.11
\citep{fiorucci2003} which allowed us to estimate extinctions to each
star individually using as many as four different colors.  We adopted
the median A$_V$, and adopted the standard deviation of the A$_V$
values as a conservative estimate of their uncertainty. For the stars
which lack reliable $V$ band photometry, we estimate their extinctions
using E(J-K$_S$).  Our extinction estimates are listed in
\autoref{tbl:member_properties}.

\subsection{H-R Diagram for Sco-Cen Members}\label{sec:hrd_km}

\begin{figure}
\begin{center}
\includegraphics[scale=0.45]{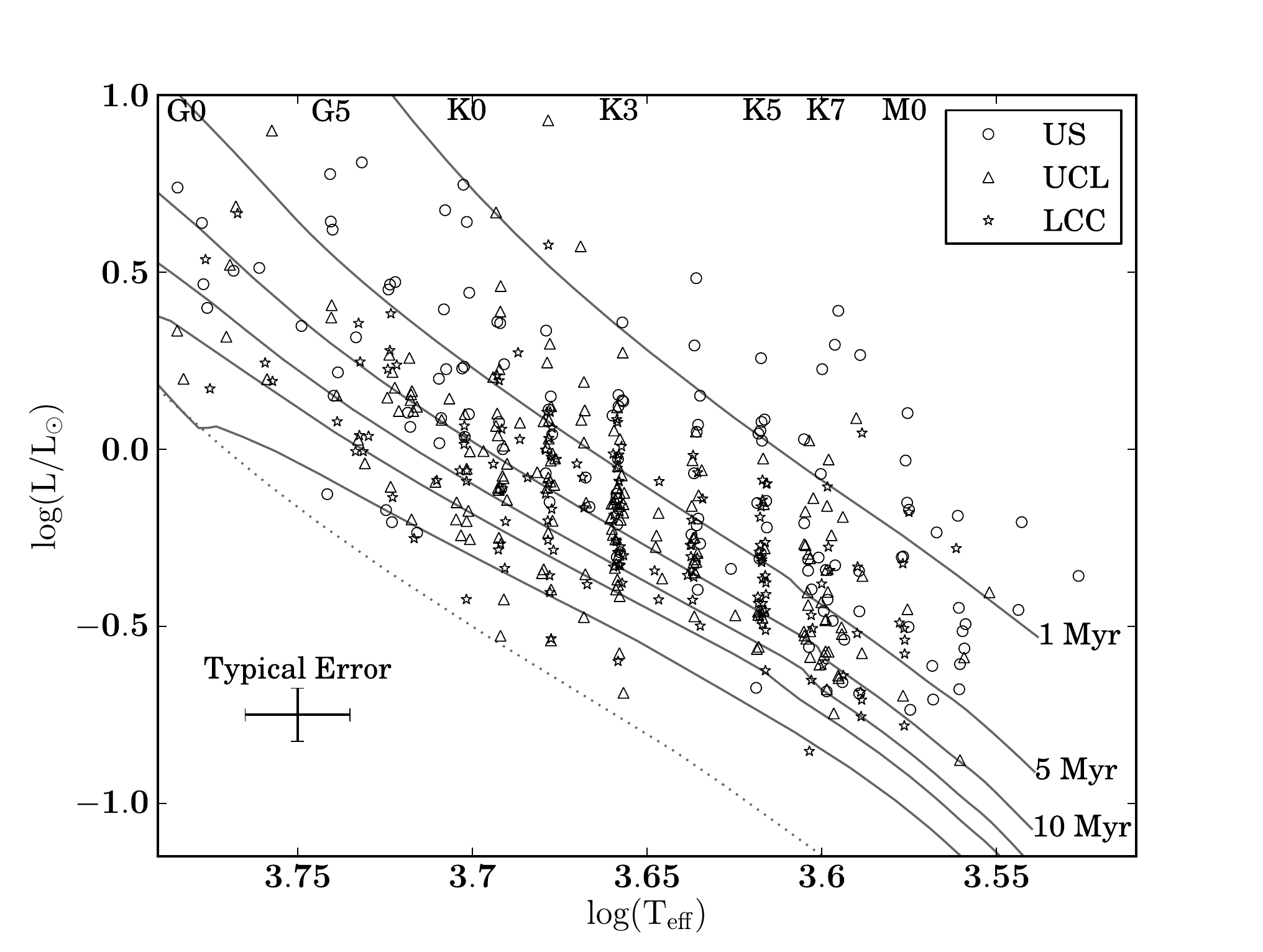}
\caption{H-R diagram for X-ray selected Sco-Cen members with
  $J$--$K_S > 0.50$ as described in the text.  Circles, triangles and
  star symbols are US, UCL and LCC candidate members, respectively.
  Plotted for comparison are 5, 10, 15, 20 and 30 Myr isochrones from
  \protect\citet{dotter2008}.  Artificial scatter of
  $\delta$\teff=$\pm$0.001~dex has been introduced for plotting
  purposes.}
\label{fig:hrd}
\end{center}
\end{figure}

We place our Sco-Cen members on a theoretical H-R diagram in order to
compare with theoretical models and obtain individual isochronal age
estimates.  We adopt the effective temperature scale (\teff) and
bolometric correction (BC$_V$) scale from the \cite{pecaut2013} 
study\footnote{
\citet{pecaut2013} adopted an absolute bolmetric magnitude for the 
Sun of ${\rm M_{\odot}=4.755~mag}$, $\sim 0.015$~mag higher than 
the recently drafted IAU bolometric magnitude scale (2015 IAU resolution 
B2).  Thus the BC scale in \citet{pecaut2013} should be shifted
down by $\sim$0.02~mag to conform to the IAU scale.  The luminosity
estimates in this present work are on the 2015 IAU scale.
},
which was constructed specifically for 5-30~Myr old pre-MS stars.
This \teff\, and BC$_V$ scale was derived by fitting the spectral
energy distributions of members of young, nearby moving groups to the
BT-Settl theoretical atmospheric models \citep{allard2012}.  Though
this \teff\, and BC scale is dependent on model atmospheres, the
method used to develop the resultant \teff\, scale is in good
agreement with \teff\, values derived from angular diameter
measurements \citep[e.g.,][]{boyajian2012b}.  We combine our
individual extinction estimates together with our kinematic parallaxes
and 2MASS J-band magnitudes to estimate the bolometric luminosities of
our stars and place them on a theoretical H-R diagram.  We compare our
data to the pre-MS models of \cite{dotter2008} in
\autoref{fig:hrd}.

\begin{figure}
\begin{center}
\includegraphics[scale=0.45]{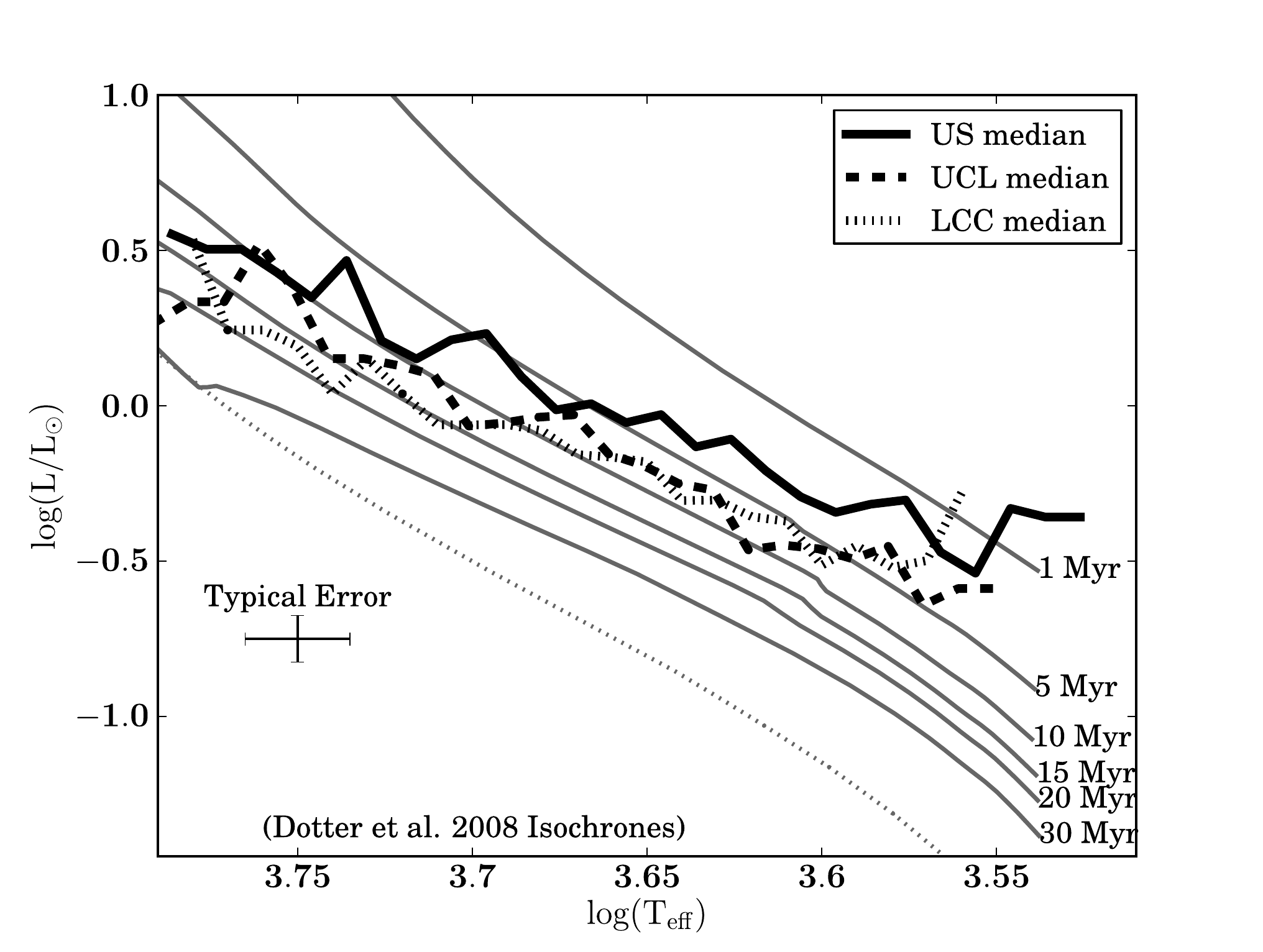}
\caption{H-R diagram with empirical isochrones for US (solid), UCL
  (dashed) and LCC (dotted).  All three subgroups exhibit a
  mass-dependent age trend.}
\label{fig:hrd_iso}
\end{center}
\end{figure}

There is considerable scatter in the H-R diagram so we construct
empirical isochrones by plotting the median luminosity along the H-R
diagram.  Though there is some scatter, the relative age rank of the
three groups is consistent with that found in \citet{mamajek2002},
\citet{preibisch2008}, \citet{pecaut2012}; from oldest to youngest:
LCC, UCL and US, with UCL and LCC approximately coeval.  The other
striking feature of the empirical isochrones is the mass-dependent age
trend.  The lower mass stars appear younger against the theoretical
isochrones than the higher mass stars.  This is the same
mass-dependent age trend seen in other studies
\citep[e.g.,][]{hillenbrand1997,hillenbrand2008,bell2012,bell2013,bell2014}.
The likely origin of the mass-dependent age trend is difficulties in 
handling convection with magnetic fields in young, low-mass stars in 
the evolutionary models, perhaps due to missing physics 
\citep[see e.g.,][]{feiden2016}.  At an age of 10~Myr and 15~Myr, our 
observational uncertainties yield individual age uncertainties of 
$\pm 3$~Myr and $\pm 4$~Myr, respectively.

\subsection{Ages}

The presence of massive stars in OB associations also allows for the
opportunity for a comparison of ages obtained through different parts
of the H-R diagram.  As massive stars burn through their nuclear fuel,
they will expand and leave the main sequence.  At the same time, the
pre-main sequence members of the OB association will be contracting
towards the main sequence.  The evolutionary models will predict ages 
for each of these segments of the H-R diagram, but different aspects 
of stellar physics are important in each of these segments.  In this 
section, we examine the ages for both the low-mass pre-MS stars as well 
as the massive main sequence turn-off stars.

\subsubsection{Pre-MS Ages}

\begin{figure}
\begin{center}
\includegraphics[scale=0.45]{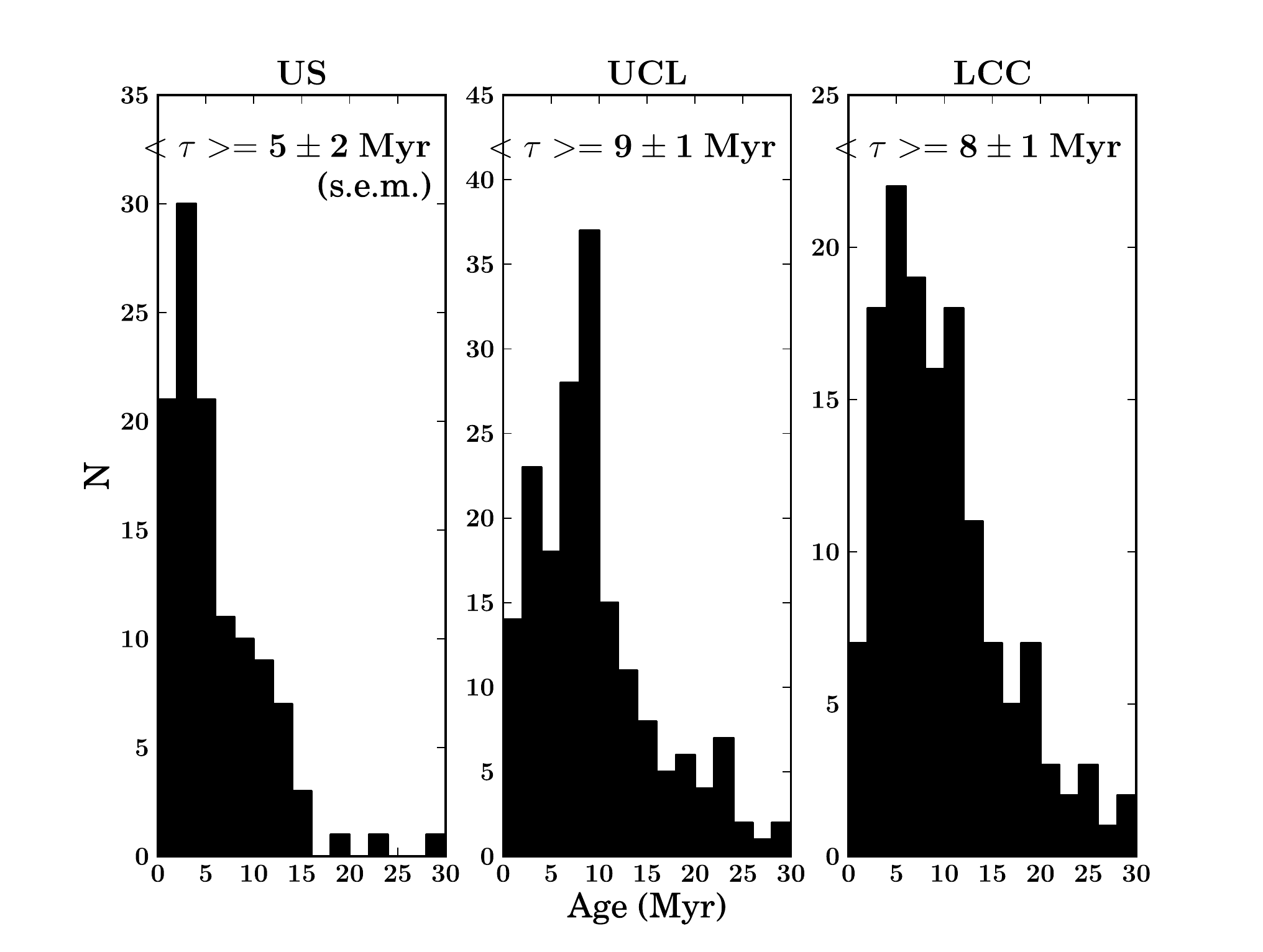}
\caption{Distribution of raw isochronal ages obtained for the Sco-Cen 
  members of our sample using the \protect\cite{dotter2008} evolutionary 
  models.}
\label{fig:agehist}
\end{center}
\end{figure}

We estimate individual ages for our pre-MS Sco-Cen members by linearly
interpolating between theoretical model isochrones.  We use the
Dartmouth models \citep{dotter2008}, the Pisa models
\citep{tognelli2011}, the PARSEC models \citep{chen2014}, and the
Exeter/Lyon models \citep{baraffe2015}.  These models each assume
slightly different stellar compositions, but are each calibrated to
reproduce the H-R diagram position of the Sun at its current age.  The
Dartmouth evolutionary models adopt the \citet{grevesse1998} solar
composition with a protosolar He fraction of Y=0.2740, while the Pisa
evolutionary models adopt the \citet{asplund2005} solar composition
with a protosolar He fraction of Y=0.2533.  The PARSEC models adopt the
\citet{caffau2008,caffau2009} solar abundances with a initial He 
fraction of Y=0.284, whereas the Exeter/Lyon models use the 
\citet{caffau2011} solar abundances, Y=0.271.  The distributions of 
ages we obtain with the Dartmouth evolutionary models is shown in
\autoref{fig:agehist}.  We obtain median ages of 5$\pm$2~Myr,
9$\pm$1~Myr and 8$\pm$1~Myr (standard error of the median) for US, UCL
and LCC, respectively.  Surprisingly, these ages are half the mean
ages obtained from the F-type members of Sco-Cen \citep{pecaut2012}.
Given the large \teff-dependent trend with age
(\autoref{fig:hrd_iso}), we decide against adopting mean subgroup
ages based on the K- and M-type members of Sco-Cen.  A detailed
discussion of our reasoning is available in 
section~\ref{sec:discuss_ages}.

In order to extract useful age estimates for the subgroups, we
re-examine the G-type pre-MS ages as well as the main-sequence
turn-off ages.  We do this for two reasons, other than the obvious
desire to quote reliable mean ages.  The first reason is that
estimating realistic intrinsic age spreads requires a reliable age as
free of systematic uncertainties as possible.  We use Monte Carlo
simulations to constrain our intrinsic age spreads, taking into
account realistic binarity statistics and observational uncertainties,
the details of which are described in section~\ref{sec:age_spreads}.
Because isochrones evenly spaced in age will not generally be evenly
spaced in luminosity, a large intrinsic spread in luminosity at a
young age could still be consistent with a small age spread, whereas
the same intrinsic spread in luminosity at an older age would imply a
larger intrinsic age spread.  The second reason we reevalute the
G-type ages and nuclear ages is that updated \teff\, and bolometric
corrections (BCs) have been made available since the ages were last
estimated in \citet{mamajek2002}.  For the hot stars, 
\citet{nieva2013} used modern non-local thermodynamic equilibrium 
(NLTE) spectral synthesis models to re-evaluate the \teff\, and BC 
scale for the massive OB stars (34,000~K $>$ \teff $>$ 15,800~K).  
Their results for dwarfs and subgiants are 1500-6000~K hotter than 
the scale of \citet{napiwotzki1993} that was adopted in the study of
\citet{mamajek2002}.  As previously mentioned, \citet{pecaut2013} have
constructed an updated \teff\, and BC scale applicable to the Sco-Cen
G-type stars.  Both of these studies should allow us to place Sco-Cen
members on the H-R diagram with reduced systematic uncertainties.

\subsubsection{Isochronal Ages for G-type Pre-MS Stars}

To estimate the ages of the G-type pre-MS stars, we collect the G-type
members from this study together with those in \citet{walter1994},
\citet{preibisch1999}, \citet{mamajek2002}, and \citet{preibisch2008}.
We supplement this with G-type stars studied in \cite{torres2006} in
the Sco-Cen field which are Li-rich and have proper motions consistent
with membership; G-type stars identified as members here conform to 
the membership criteria discussed in section~\ref{sec:membership}. 
The stellar properties for these stars are listed in
\autoref{tbl:member_properties} and \autoref{tbl:scocen_gstars},
together with the median mass and age obtained through a comparison of
their H-R diagram positions with the evolutionary models of
\cite{dotter2008}, \citet{tognelli2011}, \citet{chen2014}, and
\citet{baraffe2015}.

\begin{footnotesize}
\onecolumn
\begin{longtable}{lllllrrrrr}
\caption{Stellar parameters for Sco-Cen G-Type stars}\label{tbl:scocen_gstars} \\
\hline
Object &
2MASS &
Spectral &
Ref. &
Region &
A$_V$ &
\logt &
\logl &
Age & 
Mass \\
Name &
 &
Type &
 &
 &
(mag) &
(dex) &
(dex) &
(Myr) &
(M$_{\odot}$) \\
\hline
\endfirsthead
\multicolumn{10}{c}%
{\tablename\ \thetable\ -- \textit{Continued from previous page}} \\
\hline
Object &
2MASS &
Spectral &
Ref. &
Region &
A$_V$ &
\logt &
\logl &
Age & 
Mass \\
Name &
 &
Type &
 &
 &
(mag) &
(dex) &
(dex) &
(Myr) &
(M$_{\odot}$) \\
\hline
\endhead
\hline \multicolumn{10}{r}{\textit{Continued on next page}} \\
\endfoot
\endlastfoot
TYC 9216-0524-1 & 10495605-6951216 & G8V & T06 & LCC & 0.00$\pm$0.05 & 3.717$\pm$0.007 & -0.382$\pm$0.069 & 48 & 0.9 \\
TYC 9212-1782-1 & 10594094-6917037 & G6V & T06 & LCC & 0.00$\pm$0.02 & 3.732$\pm$0.008 & 0.428$\pm$0.085 & 6 & 1.6 \\
CD-48 6632 & 11350376-4850219 & G7V & T06 & LCC & 0.00$\pm$0.15 & 3.709$\pm$0.008 & 0.048$\pm$0.075 & 11 & 1.2 \\
HIP 57524 & 11472454-4953029 & G0V & PM & LCC & 0.18$\pm$0.04 & 3.782$\pm$0.004 & 0.413$\pm$0.061 & 17 & 1.3 \\
CPD-63 2126 & 12041439-6418516 & G8V & T06 & LCC & 0.00$\pm$0.04 & 3.717$\pm$0.007 & 0.012$\pm$0.071 & 15 & 1.2 \\
MML 3 & 12044888-6409555 & G1V & PM & LCC & 0.38$\pm$0.06 & 3.776$\pm$0.007 & 0.484$\pm$0.084 & 14 & 1.3 \\
HIP 58996 & 12054748-5100121 & G1V & PM & LCC & 0.15$\pm$0.04 & 3.776$\pm$0.007 & 0.407$\pm$0.057 & 16 & 1.3 \\
MML 7 & 12113815-7110360 & G5V & PM & LCC & 0.23$\pm$0.02 & 3.740$\pm$0.009 & 0.338$\pm$0.067 & 10 & 1.4 \\
HIP 59854 & 12162783-5008356 & G1V & PM & LCC & 0.31$\pm$0.05 & 3.776$\pm$0.007 & 0.531$\pm$0.061 & 13 & 1.4 \\
MML 14 & 12211648-5317450 & G1V & PM & LCC & 0.39$\pm$0.05 & 3.776$\pm$0.007 & 0.433$\pm$0.067 & 15 & 1.3 \\
MML 17 & 12223322-5333489 & G0V & PM & LCC & 0.24$\pm$0.03 & 3.782$\pm$0.004 & 0.448$\pm$0.066 & 16 & 1.3 \\
CD-54 4763 & 12242065-5443540 & G5V & T06 & LCC & 1.00$\pm$0.39 & 3.740$\pm$0.009 & 0.598$\pm$0.085 & 5 & 1.8 \\
HIP 60885 & 12284005-5527193 & G0V & PM & LCC & 0.20$\pm$0.04 & 3.782$\pm$0.004 & 0.482$\pm$0.057 & 15 & 1.3 \\
HIP 60913 & 12290224-6455006 & G2V & PM & LCC & 0.44$\pm$0.04 & 3.769$\pm$0.009 & 0.498$\pm$0.060 & 12 & 1.4 \\
MML 29 & 13023752-5459370 & G0V & PM & LCC & 0.44$\pm$0.10 & 3.782$\pm$0.004 & 0.257$\pm$0.072 & 23 & 1.2 \\
HIP 63847 & 13050530-6413552 & G4V & PM & LCC & 0.19$\pm$0.03 & 3.750$\pm$0.009 & 0.320$\pm$0.058 & 12 & 1.3 \\
CD-59 4629 & 13132810-6000445 & G3V & T06 & LCC & 0.10$\pm$0.04 & 3.759$\pm$0.009 & 0.289$\pm$0.068 & 16 & 1.2 \\
MML 32 & 13175694-5317562 & G0V & PM & LCC & 0.75$\pm$0.14 & 3.782$\pm$0.004 & 0.616$\pm$0.102 & 12 & 1.5 \\
MML 33 & 13220446-4503231 & G0V & PM & LCC & 0.16$\pm$0.05 & 3.782$\pm$0.004 & 0.220$\pm$0.057 & 25 & 1.2 \\
HIP 65423 & 13243512-5557242 & G3V & T06 & LCC & 0.00$\pm$0.02 & 3.759$\pm$0.009 & 0.332$\pm$0.057 & 14 & 1.3 \\
HIP 65517 & 13254783-4814577 & G1V & PM & LCC & 0.23$\pm$0.05 & 3.776$\pm$0.007 & 0.104$\pm$0.053 & 28 & 1.1 \\
MML 35 & 13342026-5240360 & G0V & PM & LCC & 0.38$\pm$0.07 & 3.782$\pm$0.004 & 0.478$\pm$0.058 & 15 & 1.3 \\
MML 37 & 13432853-5436434 & G0V & PM & LCC & 0.33$\pm$0.04 & 3.782$\pm$0.004 & 0.213$\pm$0.053 & 25 & 1.2 \\
HIP 67522 & 13500627-4050090 & G0V & PM & UCL & 0.23$\pm$0.04 & 3.782$\pm$0.004 & 0.233$\pm$0.063 & 24 & 1.2 \\
MML 42 & 14160567-6917359 & G1V & PM & LCC & 0.17$\pm$0.02 & 3.776$\pm$0.007 & 0.262$\pm$0.059 & 22 & 1.2 \\
HIP 71178 & 14332578-3432376 & G9IV & PM & UCL & 0.12$\pm$0.02 & 3.709$\pm$0.008 & 0.031$\pm$0.057 & 12 & 1.2 \\
MML 49 & 14473176-4800056 & G7IV & PM & UCL & 0.00$\pm$0.04 & 3.723$\pm$0.007 & -0.109$\pm$0.067 & 24 & 1.0 \\
MML 52 & 14571962-3612274 & G3V & PM & UCL & 0.43$\pm$0.03 & 3.759$\pm$0.009 & 0.305$\pm$0.062 & 15 & 1.3 \\
MML 56 & 15011155-4120406 & G0V & PM & UCL & 0.32$\pm$0.04 & 3.782$\pm$0.004 & 0.613$\pm$0.068 & 12 & 1.5 \\
MML 57 & 15015882-4755464 & G0V & PM & UCL & 0.24$\pm$0.04 & 3.782$\pm$0.004 & 0.333$\pm$0.071 & 20 & 1.2 \\
MML 61 & 15125018-4508044 & G2V & PM & UCL & 0.31$\pm$0.04 & 3.769$\pm$0.009 & 0.165$\pm$0.082 & 24 & 1.1 \\
MML 62 & 15180174-5317287 & G6V & PM & UCL & 0.31$\pm$0.01 & 3.732$\pm$0.008 & 0.041$\pm$0.063 & 18 & 1.1 \\
TYC 8298-1675-1 & 15193702-4759341 & G9IV & T06 & UCL & 0.00$\pm$0.02 & 3.709$\pm$0.008 & -0.336$\pm$0.069 & 38 & 0.9 \\
HIP 75483 & 15251169-4659132 & G9V & T06 & UCL & 0.00$\pm$0.16 & 3.709$\pm$0.008 & -0.220$\pm$0.067 & 27 & 1.0 \\
HIP 75924 & 15302626-3218122 & G3V & PM & UCL & 0.33$\pm$0.11 & 3.759$\pm$0.009 & 0.572$\pm$0.063 & 8 & 1.6 \\
HIP 76472 & 15370466-4009221 & G0V & PM & UCL & 0.52$\pm$0.03 & 3.782$\pm$0.004 & 0.560$\pm$0.060 & 13 & 1.4 \\
MML 69 & 15392440-2710218 & G5V & PM & UCL & 0.34$\pm$0.11 & 3.740$\pm$0.009 & 0.451$\pm$0.070 & 7 & 1.6 \\
HIP 77135 & 15445769-3411535 & G3V & PM & UCL & 0.43$\pm$0.04 & 3.759$\pm$0.009 & 0.066$\pm$0.064 & 27 & 1.1 \\
HIP 77144 & 15450184-4050310 & G0V & PM & UCL & 0.19$\pm$0.10 & 3.782$\pm$0.004 & 0.407$\pm$0.060 & 18 & 1.3 \\
HIP 77190 & 15454266-4632334 & G5IV & T06 & UCL & 0.00$\pm$0.03 & 3.740$\pm$0.009 & -0.200$\pm$0.059 & 39 & 0.9 \\
MML 72 & 15465179-4919048 & G7V & PM & UCL & 0.13$\pm$0.04 & 3.723$\pm$0.007 & 0.136$\pm$0.064 & 12 & 1.3 \\
HIP 77656 & 15511373-4218513 & G7V & PM & UCL & 0.00$\pm$0.01 & 3.723$\pm$0.007 & 0.339$\pm$0.062 & 7 & 1.6 \\
TYC 7846-1538-1 & 15532729-4216007 & G2 & T06 & UCL & 0.88$\pm$0.35 & 3.769$\pm$0.009 & 0.382$\pm$0.067 & 15 & 1.3 \\
HD 142361 & 15545986-2347181 & G2IV & W94 & US & 0.46$\pm$0.12 & 3.769$\pm$0.009 & 0.314$\pm$0.070 & 18 & 1.2 \\
HD 142506 & 15554883-2512240 & G3 & P99 & US & 0.40$\pm$0.03 & 3.759$\pm$0.009 & 0.432$\pm$0.076 & 11 & 1.4 \\
HIP 78133 & 15571468-4130205 & G3V & T06 & UCL & 0.00$\pm$0.07 & 3.759$\pm$0.009 & 0.063$\pm$0.066 & 27 & 1.1 \\
HD 143099 & 15595826-3824317 & G0V & T66 & UCL & 0.00$\pm$0.05 & 3.782$\pm$0.004 & 0.449$\pm$0.066 & 16 & 1.3 \\
CD-24 12445 & 16000078-2509423 & G0 & P99 & US & 0.37$\pm$0.17 & 3.782$\pm$0.004 & 0.256$\pm$0.074 & 23 & 1.2 \\
TYC 7333-1260-1 & 16010792-3254526 & G1V & T06 & UCL & 0.21$\pm$0.06 & 3.776$\pm$0.007 & 0.371$\pm$0.061 & 17 & 1.2 \\
MML 75 & 16010896-3320141 & G5IV & PM & UCL & 0.81$\pm$0.05 & 3.740$\pm$0.009 & 0.480$\pm$0.066 & 6 & 1.6 \\
HIP 78483 & 16011842-2652212 & G2IV & T06 & US & 0.33$\pm$0.07 & 3.769$\pm$0.009 & 0.710$\pm$0.077 & 8 & 1.7 \\
HIP 78581 & 16024415-3040013 & G1V & H82 & US & 0.08$\pm$0.05 & 3.776$\pm$0.007 & 0.353$\pm$0.069 & 18 & 1.2 \\
HIP 79462 & 16125533-2319456 & G2V & H88 & US & 0.53$\pm$0.13 & 3.769$\pm$0.009 & 0.887$\pm$0.077 & 5 & 2.0 \\
MML 82 & 16211219-4030204 & G8V & PM & UCL & 0.00$\pm$0.07 & 3.717$\pm$0.007 & 0.204$\pm$0.063 & 8 & 1.4 \\
MML 83 & 16232955-3958008 & G1V & PM & UCL & 0.48$\pm$0.12 & 3.776$\pm$0.007 & 0.212$\pm$0.084 & 24 & 1.1 \\
HIP 80320 & 16235385-2946401 & G3IV & T06 & US & 0.05$\pm$0.02 & 3.759$\pm$0.009 & 0.459$\pm$0.071 & 11 & 1.4 \\
HIP 80535 & 16262991-2741203 & G0V & H82 & US & 0.00$\pm$0.03 & 3.782$\pm$0.004 & 0.727$\pm$0.071 & 9 & 1.6 \\
HIP 80636 & 16275233-3547003 & G0V & PM & UCL & 0.36$\pm$0.05 & 3.782$\pm$0.004 & 0.563$\pm$0.060 & 13 & 1.4 \\
HIP 81380 & 16371286-3900381 & G2V & PM & UCL & 0.34$\pm$0.08 & 3.769$\pm$0.009 & 0.798$\pm$0.069 & 6 & 1.8 \\
HIP 81447 & 16380553-3401106 & G0V & PM & UCL & 0.12$\pm$0.16 & 3.782$\pm$0.004 & 0.814$\pm$0.065 & 8 & 1.7 \\
\hline
\end{longtable}
\begin{flushleft}
G-type stars not included in Table~\ref{tbl:member_properties}.
Ages and masses were estimated using the median of 
\citet{baraffe2015}, \citet{chen2014}, \citet{tognelli2011} and \citet{dotter2008}
evolutionary models.
We report the Houk types for the stars HIP~78581 (G1V), HIP~79462 (G2V) and 
HIP~80535 (G0V), but adopt the \teff\, and BC for the equivalent Gray types in our
calculations.  See \autoref{tbl:MK} for the conversions and section~\ref{sec:MK} 
for discussion. \\
Spectral Type references: 
(PM) This work;
(T06) \citet{torres2006}; 
(W94) \citet{walter1994}; 
(P99) \citet{preibisch1999};
(T66) \citet{thackeray1966};
(H82) \citet{houk1982}; 
(Ho88) \citet{houk1988} \\
Adopted median subgroup ages (Table~\ref{tbl:median_ages}) are preferable to uncertain individual ages listed above.
\end{flushleft}
\twocolumn
\end{footnotesize}

\subsubsection{Isochronal Turn-off Ages for Massive Stars}

To estimate nuclear ages, we collect membership lists for the most
massive stars from \citet{dezeeuw1999}, \citet{mamajek2002},
\citet{rizzuto2011} and \citet{pecaut2012}.  We collect Stromgren
$ubvy\beta$ and Johnson $UBV$ photometry from \citet{hauck1998} and
\citet{mermilliod1994}.  We estimate the extinction towards each early
B-type member with both the Q-method (\citealt{johnson1953}; updated
in \citealt{pecaut2013}) using Johnson $UBV$ photometry, and the
prescription of \citet{shobbrook1983}, using Stromgren $ubvy\beta$
photometry.  For all the massive stars these two methods give similar
$A_V$ within the uncertainties, so we adopt the mean value.  We
calculate the \teff\, and BC$_V$ for each star using the calibrations
with $Q$, $[u-b]$, $[c1]$ and $\beta$ in \citet{nieva2013}; we adopt
the median \teff\, when available.  We note that the calibrations
derived by \citet{nieva2013} give systematically hotter \teff\, than
the calibrations of \citet{napiwotzki1993} and \citet{balona1994},
which were used in \citet{mamajek2002} and \citet{pecaut2012},
respectively, in the most recent estimation of Sco-Cen nuclear ages.
Because of this we expect to obtain younger turn-off ages than
previously estimated.

\begin{figure}
\begin{center}
\includegraphics[scale=0.45]{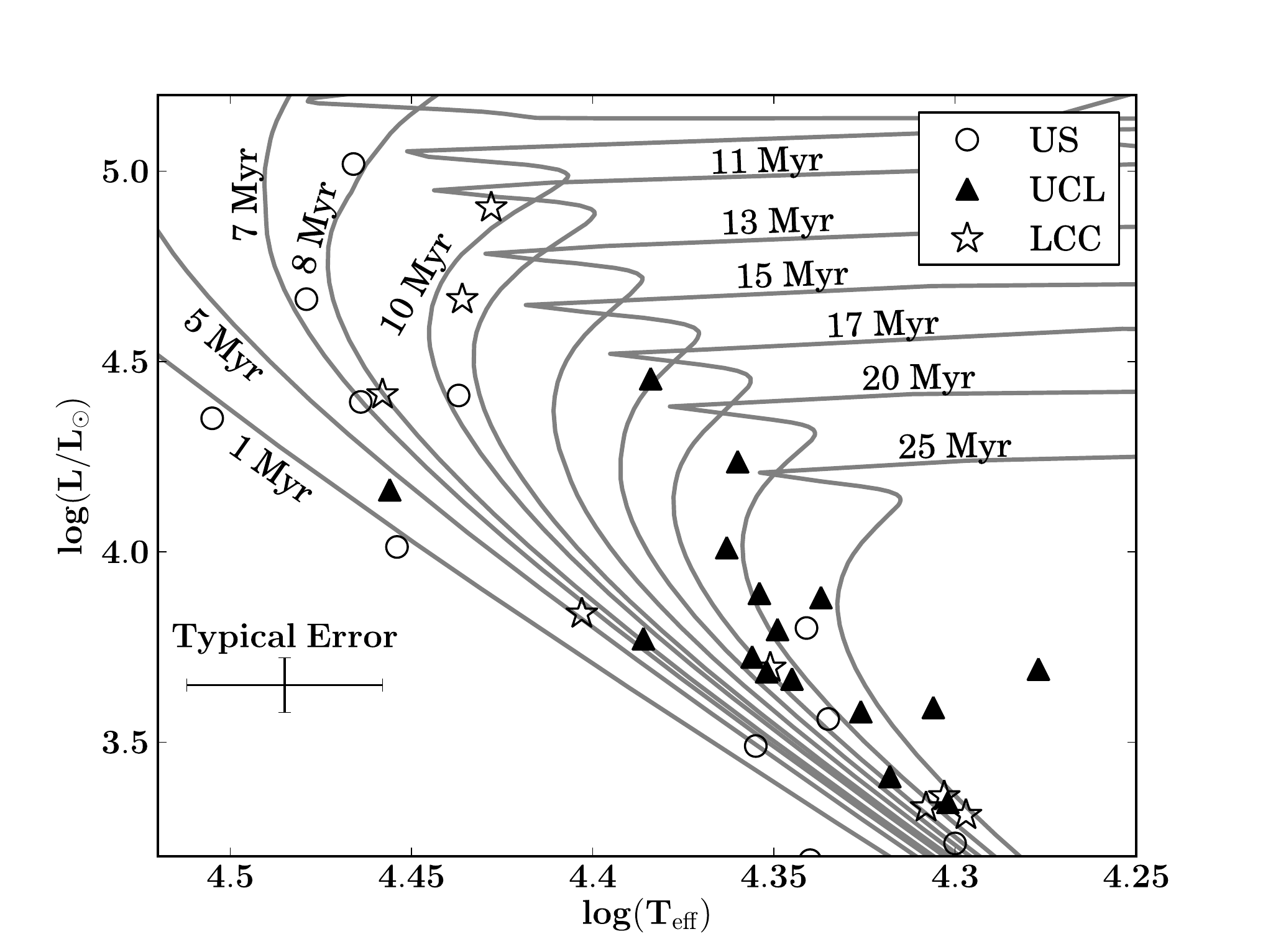}
\caption{Theoretical H-R diagram for the upper main sequence members
  of Sco-Cen with isochrones from the rotating evolutionary models of
  \protect\citet{ekstrom2012} with solar metallicity (Z = 0.014).}
\label{fig:hrd_turnoff}
\end{center}
\end{figure}

Our stellar parameters for the main-sequence turn-off in Sco-Cen are
listed in \autoref{tbl:scocen_turnoff}, along with an individual age
estimate of each object from a comparison the rotating tracks of
\citet{ekstrom2012}, rotating at $v_{eq}=0.4v_{crit}$, with the H-R
diagram position of the stars.  A plot of the main-sequence turnoff
for all three subgroups is shown in \autoref{fig:hrd_turnoff}.  To
estimate the turnoff age of Upper Sco we use $\tau$~Sco, $\omega$~Sco,
$\sigma$~Sco, $\beta^1$~Sco, $\pi$~Sco, and $\delta$~Sco.  We obtain a
median turn-off age for US of $\sim$7~Myr.  We concur with 
\citet{dezeeuw1999} that $\omicron$~Sco is unlikely to be a US 
member.\footnote{
$\omicron$~Sco is an A5II \citep{graygarrison1989} with a parallax of 
3.71$\pm$0.54~mas \citep{vanleeuwen2007}.  Its evolutionary status, 
proper motion, and distance ($\sim$270~pc) are inconsistent with 
membership in Upper Sco.} 
For Upper Centaurus-Lupus,
we use $\mu$~Cen, $\delta$~Lup, $\alpha$~Lup, $\mu^1$~Sco,
$\mu^2$~Sco, $\beta$~Lup, $\gamma$~Lup, $\nu$~Cen, $\eta$~Cen,
$\eta$~Lup, $\phi$~Cen, $\epsilon$~Lup, $\kappa$~Cen, and HR~6143.  We
obtain a median age for UCL of $\sim$19~Myr.  For Lower
Centaurus-Crux, we use $\alpha^1$~Cru, $\beta$~Cru, $\beta$~Cen,
$\delta$~Cru and $\alpha$~Mus.  We obtain a median age of
$\sim$11~Myr.  Note that these are all in the southern part of LCC,
for which \citet{preibisch2008} estimated $\sim$12~Myr.  There are no
turnoff stars in the northern part of LCC, the hottest of which 
appear to be older ($\sim$20~Myr).
Our nuclear median subgroup ages are summarized in
\autoref{tbl:median_ages}.

\onecolumn
\begin{table}
\caption{Stellar parameters for Sco-Cen main sequence turnoff stars}\label{tbl:scocen_turnoff}
\begin{tabular}{llllrrrr}
\hline
Object &
Spectral &
Ref. &
Region &
A$_V$ &
\logt &
\logl &
Age \\
Name &
Type &
 &
 &
(mag) &
(dex) &
(dex) &
Myr \\
\hline
$\tau$~Sco     & B0V       & 1 & US  & 0.109$\pm$0.029 & 4.505$\pm$0.004 & 4.351$\pm$0.125 & 1  \\
$\delta$~Sco   & B0.5IV    & 1 & US  & 0.432$\pm$0.025 & 4.479$\pm$0.030 & 4.664$\pm$0.133 & 7  \\
$\beta^1$~Sco  & B0.5V     & 1 & US  & 0.547$\pm$0.031 & 4.464$\pm$0.021 & 4.394$\pm$0.096 & 7  \\ 
$\sigma$~Sco   & B1III     & 1 & US  & 1.203$\pm$0.009 & 4.466$\pm$0.028 & 5.019$\pm$0.126 & 8  \\
$\omega$~Sco   & B1V       & 1 & US  & 0.647$\pm$0.013 & 4.454$\pm$0.016 & 4.013$\pm$0.048 & 1  \\
$\pi$~Sco      & B1V+B2:V: & 1 & US  & 0.204$\pm$0.005 & 4.437$\pm$0.030 & 4.411$\pm$0.119 & 10 \\
1~Sco          & B1.5Vn    & 1 & US  & 0.462$\pm$0.008 & 4.355$\pm$0.016 & 3.490$\pm$0.049 & 5  \\
$\nu$~Sco      & B2IV      & 1 & US  & 0.756$\pm$0.021 & 4.341$\pm$0.012 & 3.800$\pm$0.100 & 23 \\
$\beta^2$~Sco  & B2IV-V    & 1 & US  & 0.538$\pm$0.025 & 4.340$\pm$0.018 & 3.190$\pm$0.130 & $<$1 \\
$\rho$~Sco     & B2IV-V    & 1 & US  & 0.081$\pm$0.033 & 4.335$\pm$0.032 & 3.561$\pm$0.074 & 18 \\
13~Sco         & B2V       & 1 & US  & 0.112$\pm$0.005 & 4.300$\pm$0.022 & 3.234$\pm$0.052 & 15 \\
$\alpha$~Lup   & B1.5III   & 2 & UCL & 0.094$\pm$0.009 & 4.360$\pm$0.014 & 4.237$\pm$0.037 & 19 \\
$\delta$~Lup   & B1.5IV    & 2 & UCL & 0.030$\pm$0.017 & 4.384$\pm$0.033 & 4.455$\pm$0.145 & 15 \\
$\mu^1$~Sco    & B1.5IV    & 2 & UCL & 0.060$\pm$0.010 & 4.363$\pm$0.030 & 4.011$\pm$0.142 & 19 \\
$\eta$~Cen     & B1.5Vn    & 2 & UCL & 0.021$\pm$0.021 & 4.349$\pm$0.021 & 3.796$\pm$0.024 & 21 \\
$\beta$~Lup    & B2III     & 2 & UCL & 0.045$\pm$0.008 & 4.354$\pm$0.018 & 3.891$\pm$0.043 & 20 \\
HR~6143        & B2III     & 2 & UCL & 0.161$\pm$0.016 & 4.326$\pm$0.019 & 3.580$\pm$0.050 & 23 \\
$\nu$~Cen      & B2IV      & 2 & UCL & 0.025$\pm$0.003 & 4.386$\pm$0.031 & 3.772$\pm$0.070 & 8  \\
$\mu^2$~Sco    & B2IV      & 2 & UCL & 0.051$\pm$0.014 & 4.356$\pm$0.033 & 3.724$\pm$0.072 & 17 \\
$\phi$~Cen     & B2IV      & 2 & UCL & 0.000$\pm$0.019 & 4.345$\pm$0.030 & 3.666$\pm$0.069 & 19 \\
$\gamma$~Lup   & B2IV      & 2 & UCL & 0.018$\pm$0.003 & 4.337$\pm$0.030 & 3.880$\pm$0.085 & 24 \\
$\kappa$~Cen   & B2IV      & 2 & UCL & 0.015$\pm$0.008 & 4.306$\pm$0.029 & 3.591$\pm$0.082 & 30 \\
$\upsilon^1$~Cen & B2IV-V  & 2 & UCL & 0.009$\pm$0.020 & 4.318$\pm$0.027 & 3.410$\pm$0.062 & 18 \\
$\epsilon$~Lup & B2IV-V    & 2 & UCL & 0.035$\pm$0.022 & 4.277$\pm$0.029 & 3.692$\pm$0.114 & $>30$ \\
$\mu$~Cen      & B2IV-Ve   & 2 & UCL & 0.257$\pm$0.030 & 4.456$\pm$0.040 & 4.163$\pm$0.092 & 4  \\
$\chi$~Cen     & B2V       & 2 & UCL & 0.018$\pm$0.012 & 4.302$\pm$0.025 & 3.342$\pm$0.061 & 23 \\
$\eta$~Lup     & B2.5IV    & 2 & UCL & 0.003$\pm$0.003 & 4.352$\pm$0.031 & 3.685$\pm$0.070 & 17 \\
$\tau$~Lib     & B2.5V     & 2 & UCL & 0.028$\pm$0.006 & 4.242$\pm$0.031 & 3.218$\pm$0.071 & $>30$ \\
$\theta$~Lup   & B2.5Vn    & 2 & UCL & 0.031$\pm$0.020 & 4.248$\pm$0.035 & 3.106$\pm$0.092 & $>30$ \\
HR~5471        & B3V       & 2 & UCL & 0.045$\pm$0.002 & 4.251$\pm$0.029 & 3.034$\pm$0.064 & $>30$ \\
HR~5378        & B7IIIp    & 2 & UCL & 0.038$\pm$0.012 & 4.273$\pm$0.019 & 3.152$\pm$0.049 & 28 \\
$\beta$~Cru    & B0.5III   & 2 & LCC & 0.077$\pm$0.006 & 4.458$\pm$0.014 & 4.414$\pm$0.079 & 8  \\
$\alpha^1$~Cru & B0.5IV    & 2 & LCC & 0.018$\pm$0.041 & 4.436$\pm$0.038 & 4.664$\pm$0.094 & 11 \\
$\beta$~Cen    & B1III     & 2 & LCC & 0.075$\pm$0.031 & 4.428$\pm$0.022 & 4.905$\pm$0.072 & 11 \\
$\xi^2$~Cen    & B1.5V     & 2 & LCC & 0.061$\pm$0.005 & 4.308$\pm$0.029 & 3.329$\pm$0.069 & 18 \\
HR~4618        & B2IIIne   & 2 & LCC & 0.043$\pm$0.009 & 4.247$\pm$0.010 & 3.058$\pm$0.075 & $>30$ \\
$\delta$~Cru   & B2IV      & 2 & LCC & 0.000$\pm$0.007 & 4.403$\pm$0.025 & 3.838$\pm$0.056 & 6  \\
$\alpha$~Mus   & B2IV-V    & 2 & LCC & 0.015$\pm$0.015 & 4.351$\pm$0.024 & 3.696$\pm$0.054 & 18 \\
$\mu^1$~Cru    & B2IV-V    & 2 & LCC & 0.067$\pm$0.009 & 4.297$\pm$0.025 & 3.309$\pm$0.057 & 24 \\
$\sigma$~Cen   & B2V       & 2 & LCC & 0.060$\pm$0.022 & 4.303$\pm$0.032 & 3.358$\pm$0.073 & 23 \\
$\zeta$~Cru    & B2.5V     & 2 & LCC & 0.021$\pm$0.021 & 4.246$\pm$0.025 & 3.048$\pm$0.070 & $>30$ \\
\hline
\end{tabular}
\begin{flushleft}
Ages were estimated using the \citet{ekstrom2012} rotating evolutionary models with solar abundances.  
Spectral Type references: (1) \citet{hiltner1969}; (2) \citet{garrison1967};
\end{flushleft}
\end{table}
\twocolumn

We summarize our derived nuclear, F-type pre-MS and G-type pre-MS ages
in \autoref{tbl:median_ages} with our adopted values for each
subregion.

\begin{table}
\caption{Adopted subgroup ages}\label{tbl:median_ages}
\begin{tabular}{lrrr}
\hline
Method &
US &
UCL &
LCC \\
 &
(Myr) &
(Myr) &
(Myr) \\
\hline
MS Turnoff    &  7$\pm$2 & 19$\pm$2 & 11$\pm$2 \\
G-Type Pre-MS & 10$\pm$1 & 15$\pm$1 & 15$\pm$1 \\
F-Type Pre-MS$^a$ & 13$\pm$1 & 16$\pm$1 & 17$\pm$1 \\
\hline 
Adopted       & 10$\pm$3 & 16$\pm$2 & 15$\pm$3 \\
\hline 
\end{tabular}
\begin{flushleft}Uncertainties reported above are the standard
error of the mean which represents the uncertainty in how well 
the mean value is characterized; these numbers do not 
represent the spread in ages.  
Median ages are derived considering the
B-type main-sequence turn-off, the F-type pre-MS turn-on 
and the pre-MS G-type stars in each subgroup. \\
$^a$ Adopted F-type Pre-MS ages from \citet{pecaut2012}.
\end{flushleft}
\end{table}

\subsection{Intrinsic Age Spreads \label{sec:age_spreads}}

The H-R diagram positions of Sco-Cen members have a large degree of
scatter, and hence a large apparent scatter in inferred ages.  Some of
this scatter is due to observational uncertainties and unresolved
multiplicity, but some scatter may be due to true age spreads within
the subgroups.  Previous studies have found very small intrinsic age
spreads in US, but larger age spreads in UCL and LCC
\citep{preibisch2002,slesnick2006,mamajek2002,preibisch2008,pecaut2013}.
We perform Monte Carlo simulations in order to model the scatter
caused by the observational uncertainties, the effects of unresolved
binarity, and an intrinsic age spread.  We create many populations of
10$^4$ simulated members with a gaussian distribution of ages,
assuming a \cite{kroupa2001b} initial mass function (IMF), a spot 
filling factor from 0\% to 50\%, a multiplicity fraction of 0.44 and 
companion mass ratio distribution given by a flat power law distribution 
with $\gamma$=0.3 (using multiplicity properties for 0.7\msun $\simless$ 
M$_* \simless$ 1.3\msun\, population~I main sequence stars as summarized 
in \citealt{duchene2013} and \citealt{raghavan2010}).  To create the 
simulated population, each star is assigned a mass and an age, 
determined to be either binary or single, assigned a companion mass if 
binary, and assigned a spot coverage ratio from a random uniform 
distribution ranging from 0\% to 50\%, which alters its H-R diagram 
position according to the \cite{somers2015b} correction factors.
We then introduce dispersion in their H-R diagram positions 
with the median observational uncertainties from our sample.  This is 
designed to simulate a population with a given mean age, an intrinsic 
age spread, the effects of unresolved binarity, cool spots, and 
observational uncertainties.  Following \citet{hillenbrand2008}, we 
compare the the observed luminosity spreads around the empirical 
isochrones in Sco-Cen members to the simulated luminosity spreads.  
We emphasize that in
this comparison, we only compare the distributions of luminosities
around the median.  We do not compare the median ages obtained since
we observe a mass-dependent age trend. However, unlike
\citet{hillenbrand2008}, we only model star formation with a given
mean age and intrinsic (gaussian) spread, rather than consider
accelerating star formation or other distributions of ages.  

\begin{figure}
\begin{center}
\includegraphics[scale=0.45]{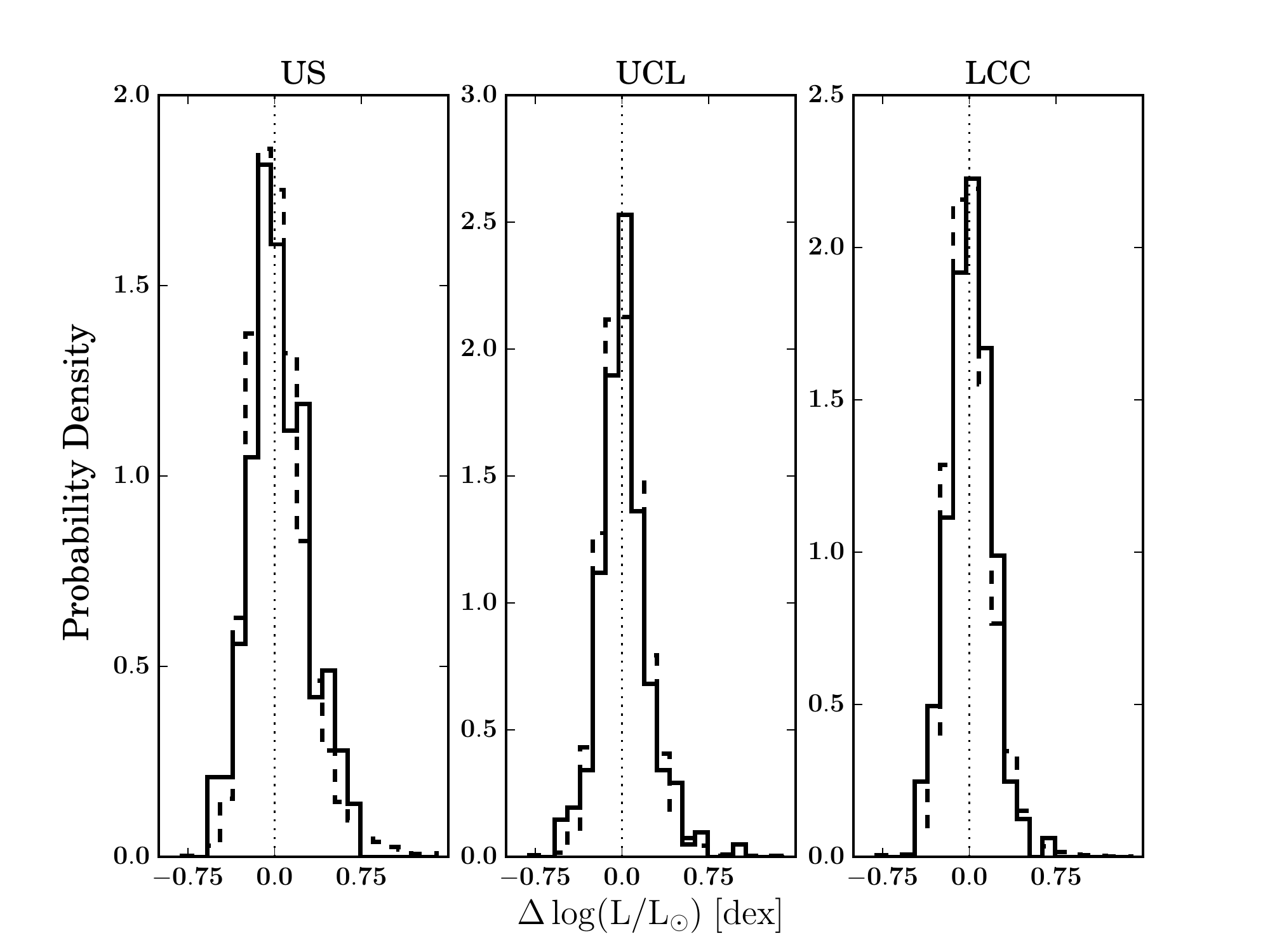}
\caption{Observed luminosity spreads around the empirical isochrone
  (solid line) compared with the same for a simulated population
  (dashed line).  The simulations used mean ages of 10~Myr, 16~Myr and
  15~Myr for US, UCL, and LCC, respectively.  We obtained best-fit
  intrinsic age spreads of $\pm$7~Myr, $\pm$7~Myr, $\pm$6~Myr for
  US, UCL and LCC, shown above with simulations made with the
  \protect\citet{baraffe2015} evolutionary models.}
\label{fig:luminositydist}
\end{center}
\end{figure}

We compare our observed luminosity spread with the luminosity spread
from the simulations using a Anderson-Darling goodness-of-fit test
\citep[e.g.,][]{hou2009}, and adopt the age spread which best matches the 
simulated luminosity spreads for each Sco-Cen subgroup.  We adopted 
the median subgroup ages listed in \autoref{tbl:median_ages} as the 
mean age of our simulated populations to compare the luminosity 
spreads in our simulations to those from our observations.  We adopt 
1$\sigma$ gaussian age spreads of $\pm$7~Myr, $\pm$7~Myr and 
$\pm$6~Myr for US, UCL and LCC, respectively.  A plot showing the 
best matches to the observed luminosity spreads is shown in
\autoref{fig:luminositydist}.  Our results are summarized in
\autoref{tbl:age_spreads}.  These age spreads indicate that 68\%
of star formation in each subgroup occured over a period of 
$\sim$14~Myr for US and UCL, and over a period of $\sim$12~Myr for 
LCC.

\begin{table}
\caption{Subregion intrinsic age spreads}\label{tbl:age_spreads}
\begin{tabular}{ @{\hspace{0mm}}l @{\hspace{1.0mm}}l @{\hspace{1.0mm}}l @{\hspace{1.0mm}}l }
\hline
Evolutionary &
US at 10~Myr &
UCL at 16~Myr &
LCC at 15~Myr \\
Models &
(Myr) &
(Myr) &
(Myr) \\
\hline 
Exeter/Lyon      & $\pm$8 & $\pm$8 & $\pm$6 \\
Parsec           & $\pm$6 & $\pm$7 & $\pm$6 \\
Pisa             & $\pm$8 & $\pm$8 & $\pm$6 \\
Dartmouth        & $\pm$6 & $\pm$7 & $\pm$5 \\
\hline 
Adopted Intrinsic Age Spreads & $\pm$7 & $\pm$7 & $\pm$6 \\
\hline 
\end{tabular}
\begin{flushleft}
Age spreads were estimated comparing observed luminosity 
spreads to simulated populations with a given intrinsic age 
spread, taking into account multiplicity, spots and observational 
uncertainties.  Adopted median ages are listed in 
\autoref{tbl:median_ages}.  
\end{flushleft}
\end{table}

\subsection{Spatial Variation of Ages}\label{sec:age_map}

\cite{blaauw1964} divided Sco-Cen into the subgroups Upper Scorpius,
Upper Centaurus-Lupus and Lower Centaurus Crux.  However,
\cite{rizzuto2011} notes that, based on their updated membership study
of Sco-Cen, the distribution of probable members indicates that the
subgroups cannot be defined in a non-arbitrary manner.  US has
consistently been shown to be younger than the other two subgroups,
and UCL and LCC have typically been assigned similar ages
\citep{preibisch2008,mamajek2002,pecaut2012}.  Given the large spatial
extent of UCL and LCC, however, it is too simplistic to place them
into two groups, each characterized by a single mean age.  For
example, the $\sim50$~pc size of LCC and adopted mean age of 
$\sim17$~Myr leads to an expected individual age uncertainty of 
$\sim$50\%, or $\sim$8~Myr, based on a star formation 
timescale-size relation \citep[see discussion in][]{soderblom2014}.  
Here we attempt to examine the age structure of the entire association 
to discern if we see evidence for spatial substructure based on
systematic differences in age as a function of position on the sky.

One possible method to investigate spatial variations of ages would be 
to simply evaluate the age of each member against pre-MS evolutionary tracks
and spatially average their ages to create an age map.  However, given 
the strong mass-dependent age trend we find with all the evolutionary 
tracks, this would tend to show regions with large numbers of low-mass 
stars as younger regions, when in fact this may be a systematic effect 
due to the evolutionary tracks.  Therefore we do not use 
this method.  

\begin{figure}
\begin{center}
\includegraphics[scale=0.45]{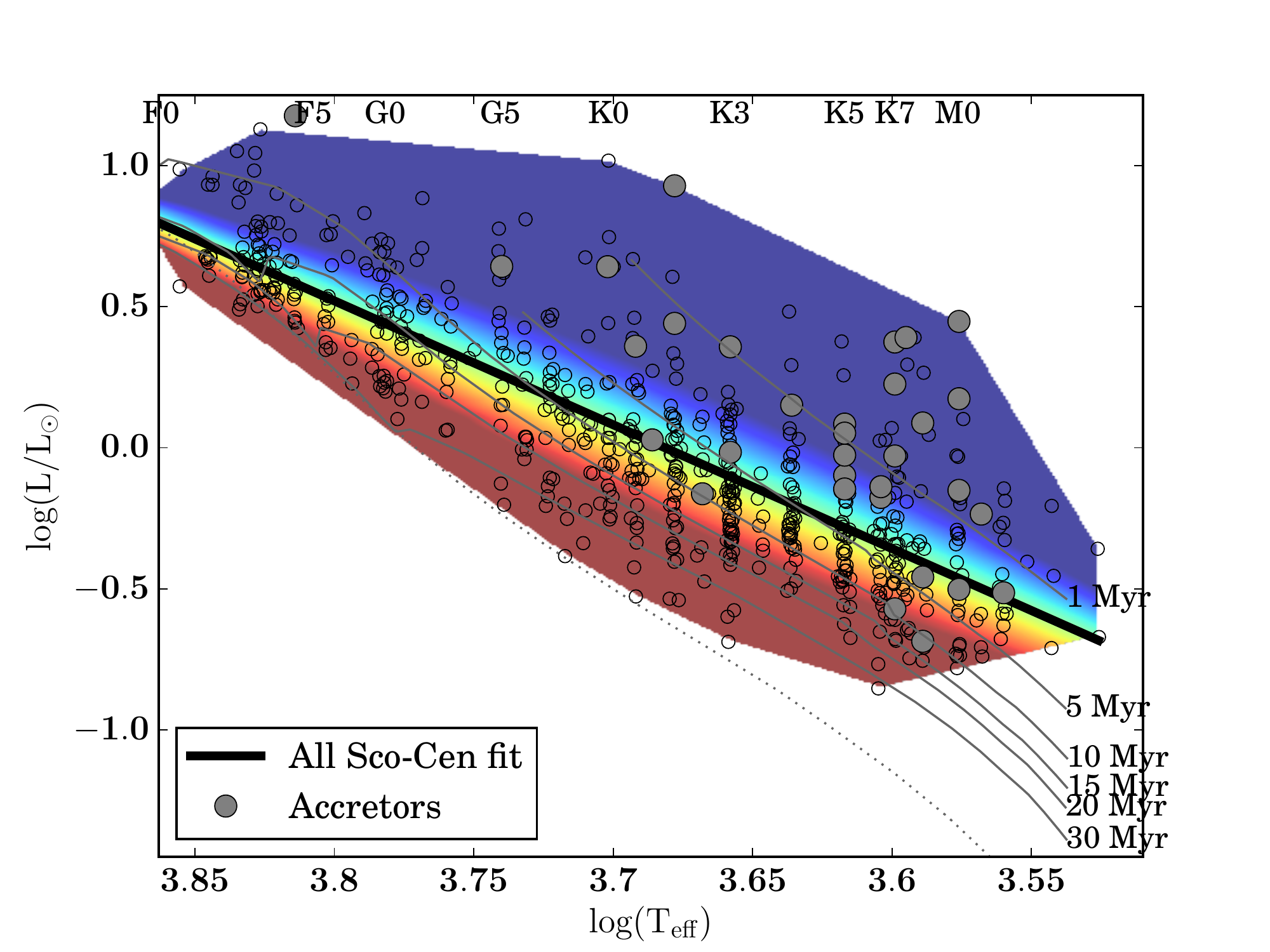}
\caption{Empirical isochrone for 657 F/G/K-type members of all
  subgroups of Sco-Cen.  We fit a line to create the empirical
  isochrone.  We use individual stars' offsets above or below this 
  empirical isochrone to create a relative age map, shown in
  \autoref{fig:agemap}.  The colored regions shown here correspond
  to the ages on the map in \autoref{fig:agemap}.
}
\label{fig:agemaphrd}
\end{center}
\end{figure}

Systematically younger regions will tend to be more luminous than the 
mean association luminosity as a function of \teff.  This is 
comparison does not depend on {\it any} theoretical models.
To probe for statistically significant spatial variations of ages,
{\it independent} of any evolutionary tracks, we look for concentrations
of stars which lie above or below the average luminosity as a function of
\teff.  We use all the pre-MS members which have been studied 
spectroscopically.  We place all stars from all three subgroups together 
on the H-R diagram and construct an empirical isochrone for the entire
association, by fitting a line to \logl\, as a function of \teff.  This is 
plotted in \autoref{fig:agemaphrd}.  Each star lies on the H-R diagram 
above or below this empirical isochrone.  For each star we calculate 
this offset in units of the luminosity spread $\sigma$ above or below 
the linear fit:

offset = $($\logl$ - <$\logl$>) / \sigma_{\log({\rm L/L_{\odot}})}$

\begin{landscape}
\begin{figure}
\includegraphics[scale=0.67]{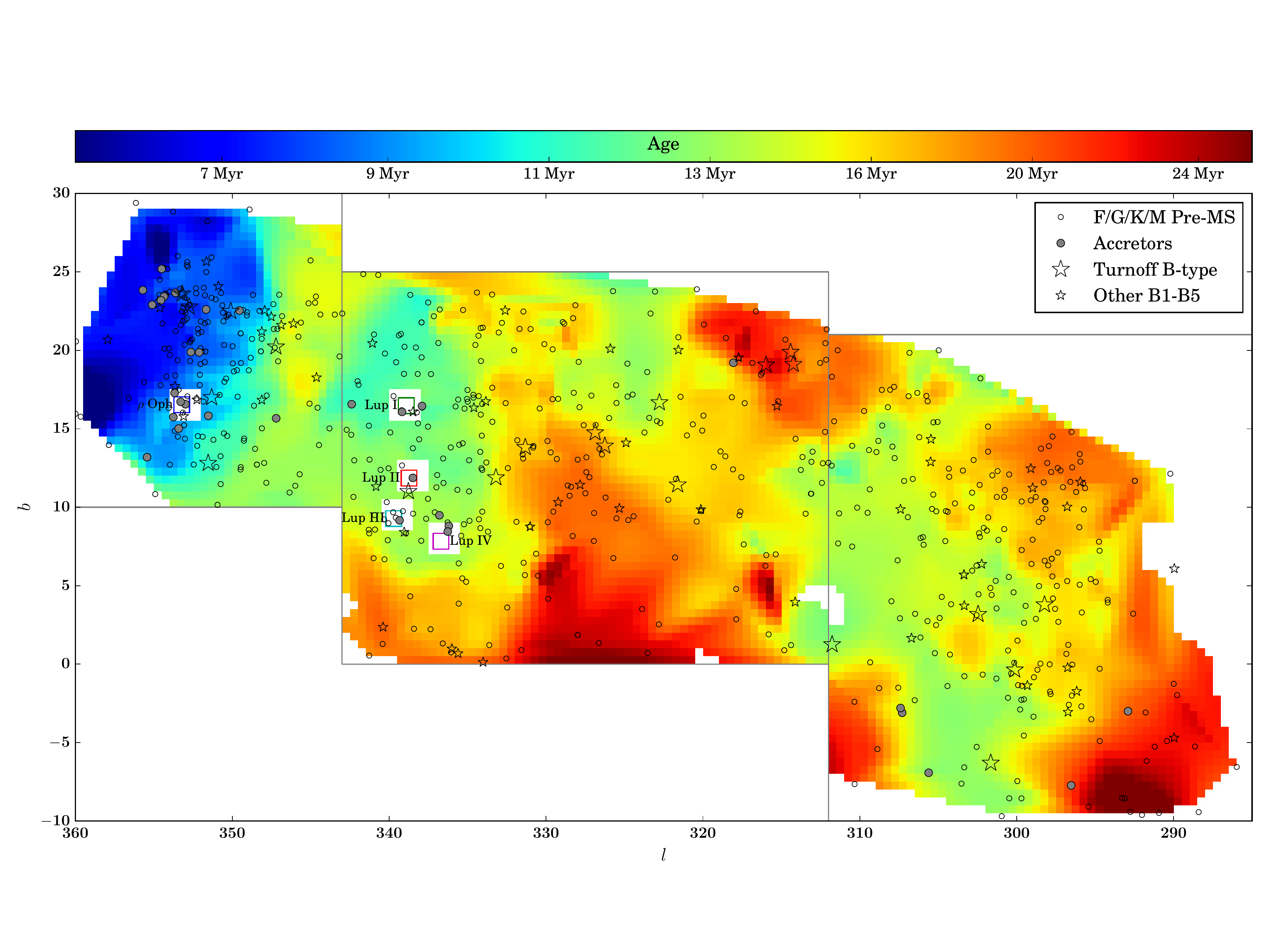}
\caption{Spatial distribution of 657 F/G/K/M-type pre-MS members
  (solid dots) with spatially averaged median ages plotted on a 
  ``age map''.  The map is created by evaluating the median 
  luminosity offset relative to the spread in luminosity 
  $($\logl$ - <$\logl$>) / \sigma_{\log({\rm L/L_{\odot}})}$ 
  of stars in a 5$^{\circ}$ radius each $(l, b)$.
  These offsets are then correlated with ages using the F5 through G9
  stars that fall in those offset bins.  We have masked the age map
  over regions cospatial with the Ophiuchus and Lupus clouds.}
\label{fig:agemap}
\end{figure}
\end{landscape}

We then spatially average their {\it offset}, above or below the empirical
isochrone, in galactic coordinates, shown in \autoref{fig:agemap}.  
Regions on the sky with stars which are systematically more luminious than 
the average of the association will lie systematically above the empirical 
isochrone and will appear on the age map as younger.  Likewise, stars in 
older regions will tend to be less luminous on average than the average of 
the association.  In \autoref{fig:agemap}, we plot a spatial intensity map 
of the median offset from the empirical isochrone.  The colored regions in 
\autoref{fig:agemap} correspond to the same colored regions above
or below the empirical isochrone in \autoref{fig:agemaphrd}.  Spatially 
averaging their offset from the empirical isochrone allows us to determine 
which regions have concentrations of systematically younger or older stars, 
relative to the age of the entire association, and independent of any 
theoretical models.  This method avoids biases due to any mass-dependent 
age trends, and can identify younger or older regions without reference to 
any evolutionary tracks.  The median distance of the F- and G-type stars is
within $\sim$4~pc of the median distance of the K-type stars in each 
subgroup, and is not systematically biased nearer or farther.  Furthermore, 
the variable extinction across the association ranges from $\sim 0.0-0.5$~mag 
(25\% to 75\% interquartile range), which would not significantly bias the
spatial distribution of ages.  Thus it is likely that our selection methods 
do not exhibit any systematic spatial biases for younger or older ages 
across the association that would bias the age map.  

We wish to assign mean isochronal ages of these regions using evolutionary 
models.  In order to assign ages to the regions shown on the map, we use 
the F5 through G9 stars to anchor the regions to an age.  For example, the 
F5 through G9 stars that lie $0.33\sigma$ to $0.66\sigma$ {\it below} the 
median isochrone, irrespective of their location in the association, have 
a median age of 18~Myr.  Therefore, any region with a median offset of 
$0.33\sigma$ to $0.66\sigma$ below the empirical isochrone is assigned an 
age of $\sim$18~Myr.  The age map makes use of 657 F-M pre-MS stars to 
establish {\it relative} ages from the mean Sco-Cen empirical isochrone, 
but the ages are adopted from the F5-G9 stars.  We discuss our motivations 
for adopting the F- and G-type ages in section~\ref{sec:discuss_ages}. 

\clearpage

\subsection{Circumstellar Disks}

\subsubsection{Spectroscopic Accretion Disk Fraction}

Giant planet formation is fundamentally limited by the lifetimes of
gas-rich protoplanetary disks surrounding the host star 
\citep{pollack1996}.  The gas disk dissipation timescale therefore 
provides an upper limit to the giant planet formation timescale.  
Differences in disk dissipation timescales for stars in different 
mass bins can provides critical data for inferring how the planet 
formation process differs around stars of
various masses.  Additionally, a census for gas-rich circumstellar
disks allows for follow--up studies of the gas disk itself
\citep[e.g.,][]{zhang2013} or the star--disk interaction
\citep[e.g.,][]{bouvier2007,nguyen2009}.

\begin{figure}
\begin{center}
\includegraphics[scale=0.45]{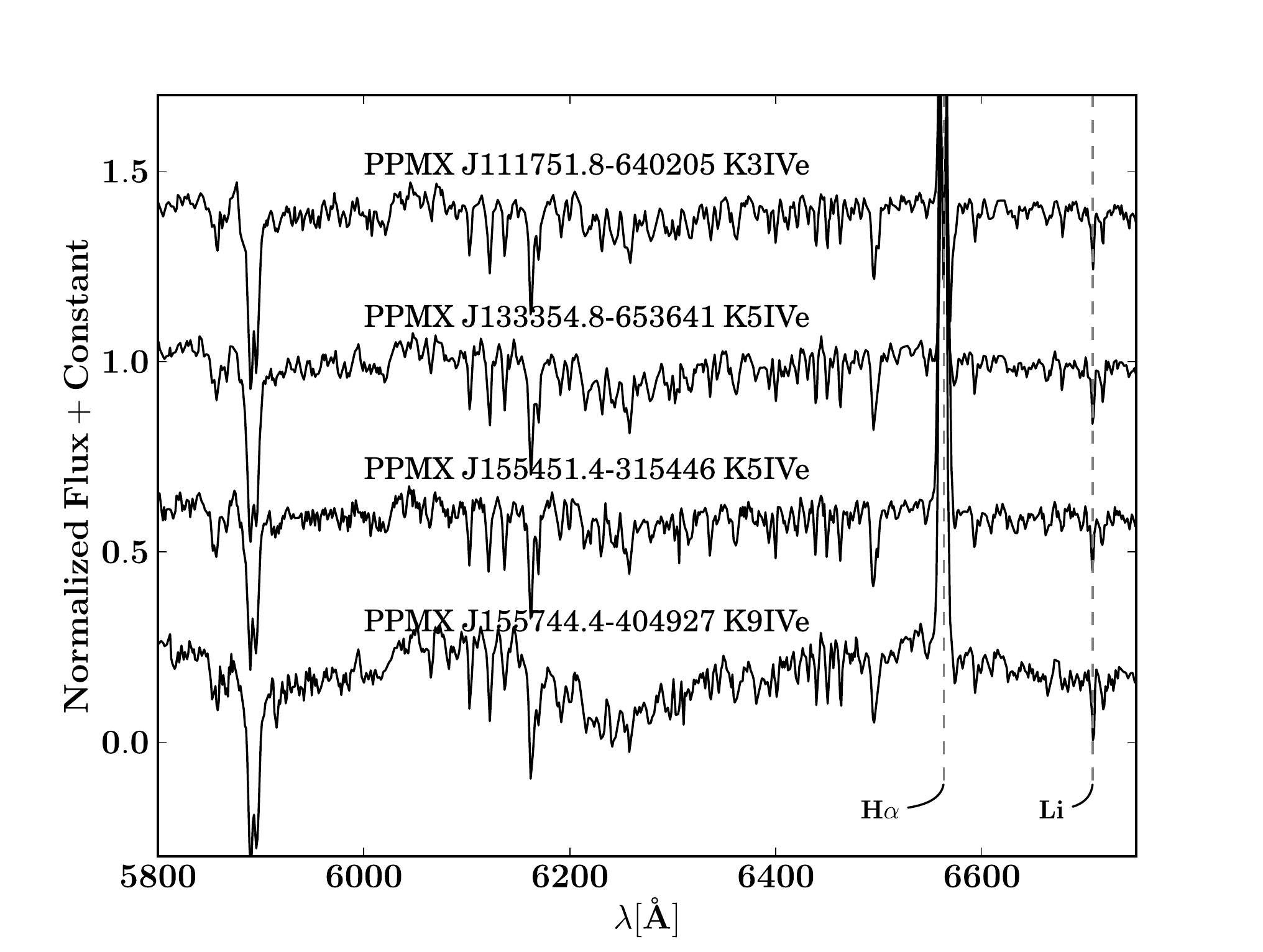}
\caption{Representative spectra of new Sco-Cen members with strong
  H$\alpha$ emission consistent with accretion.}
\label{fig:spectra}
\end{center}
\end{figure}

Here we perform a census of accretion disks for our sample using
H$\alpha$ emission as an accretion diagnostic.  Various criterion have
been proposed using H$\alpha$ as an accretion indicator, and we adopt
the spectral type dependent empirical criterion of \cite{barrado2003}.
If our measured or adopted EW(H$\alpha$) (see
\autoref{tbl:member_properties} ) exceeds the \cite{barrado2003}
criterion, we count the object as an accretor.  Our accretion disk
fraction excludes the 28 Sco-Cen members in our sample which lack
H$\alpha$ measurements.  Sample spectra for Sco-Cen members with
H$\alpha$ in emission consistent with accretion are shown in
\autoref{fig:spectra}.

\begin{figure}
\begin{center}
\includegraphics[scale=0.45]{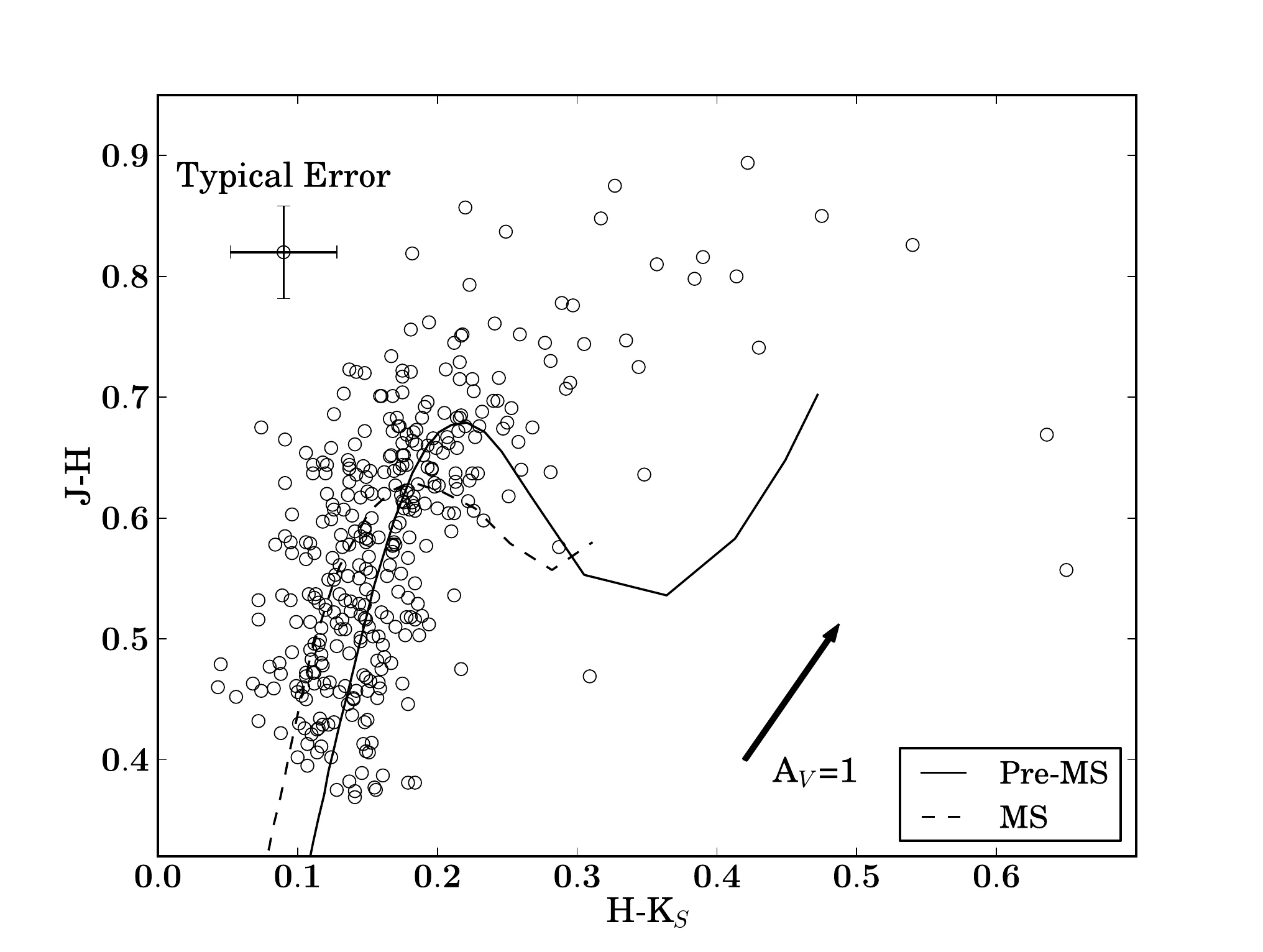}
\caption{$H$--$K_S$ versus $J$--$H$ colors for new candidate Sco-Cen
  members.  The dwarf and pre-MS stellar locus from
  \protect\cite{pecaut2013} are included shown as the dashed and solid
  lines, respectively.}
\label{fig:hk_jh}
\end{center}
\end{figure}

\begin{figure}
\begin{center}
\includegraphics[scale=0.45]{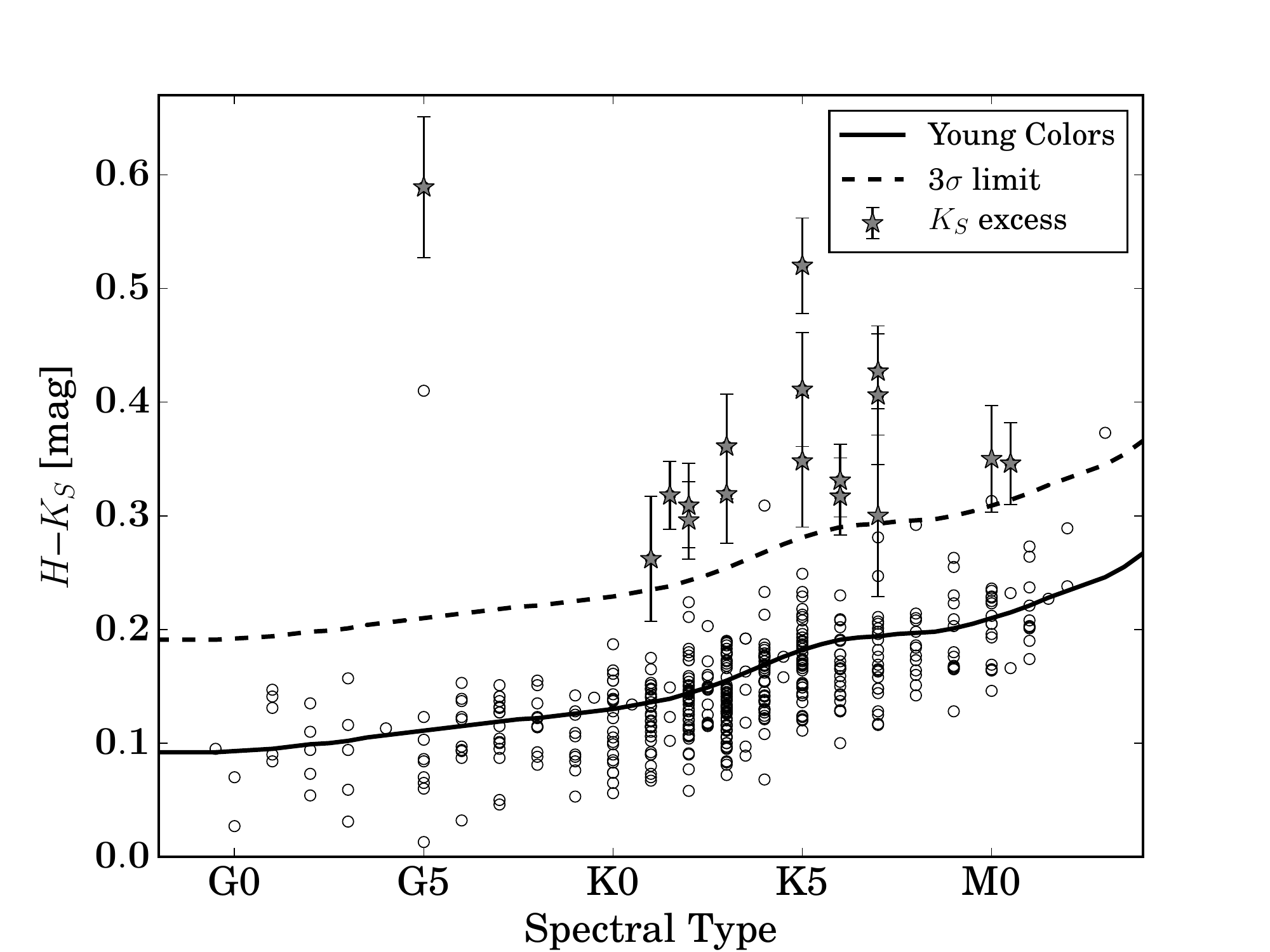}
\caption{Spectral type versus dereddened $H$--$K_S$ color for Sco-Cen
  members.  The solid line is the pre-MS photospheric colors from
  \protect\cite{pecaut2013}.  The dashed line is the 3$\sigma$
  dispersion in the photospheric colors.  Members with color excesses
  above the 3$\sigma$ dispersion in the photospheric colors in this
  and all bands at longer wavelength are identified as having a $K_S$
  band excess (gray stars).}
\label{fig:spt_hk}
\end{center}
\end{figure}

Using the \citet{barrado2003} criteria, we identified 10/108 accretors
in US, or a spectroscopic accretion disk fraction of
9.3$^{+3.6}_{-2.1}$\%.  In UCL and LCC we have 5/154 and 4/127 
accretors, or spectroscopic accretion disk fractions of 
3.2$^{+2.1}_{-0.9}$\% and 3.1$^{+2.4}_{-0.9}$\%, respectively.
These disk fractions include all K and M-type members in our sample 
with EW(H$\alpha$) measurements, and therefore have masses 
predominantly $\sim$0.7-1.3\msun, with a few M-type stars as low as 
$\sim$0.5\msun; for statistics on the K-type members {\it only}, refer 
to \autoref{tbl:excess_stats}.

\begin{table*}
\caption{Infrared excess and spectroscopic accretion disk fractions for K-Type ($\sim$0.7-1.3\msun) members in Sco-Cen}\label{tbl:excess_stats}
\begin{tabular}{lrrr}
\hline
Band/Criteria/Disk Type &
US &
UCL &
LCC \\
\hline
EW(H$\alpha$) & 6/84  (7.1$^{+3.9}_{-1.9}\%$)  & 5/145  (3.4$^{+2.2}_{-1.0}\%$)  & 4/119 (3.4$^{+2.5}_{-1.0}\%$) \\
$K_S$         & 6/89  (6.7$^{+3.7}_{-1.8}\%$)  & 6/158  (3.8$^{+2.2}_{-1.0}\%$)  & 2/119 (1.7$^{+2.1}_{-0.5}\%$) \\
$W1$          & 11/89 (12.4$^{+4.3}_{-2.7}\%$) & 9/157  (5.7$^{+2.4}_{-1.3}\%$)  & 4/119 (3.4$^{+2.5}_{-1.0}$\%) \\
$W2$          & 14/89 (15.7$^{+4.6}_{-3.1}\%$) & 9/157  (5.7$^{+2.4}_{-1.3}\%$)  & 4/119 (3.4$^{+2.5}_{-1.0}$\%) \\
$W3$          & 24/89 (27.0$^{+5.2}_{-4.1}\%$) & 17/157 (10.8$^{+3.0}_{-2.0}\%$)  & 8/118 (6.8$^{+3.1}_{-1.6}\%$) \\
$W4$          & 25/36 (69.4$^{+6.5}_{-8.5}\%$) & 33/72  (45.8$^{+5.9}_{-5.6}\%$) & 8/86  (9.3$^{+4.1}_{-2.2}\%$) \\
\hline
Full          & 8/89 (9.0$^{+4.0}_{-2.2}$\%) & 8/157 (5.1$^{+2.4}_{-1.2}$\%) & 4/118 (3.4$^{+2.5}_{-1.0}$\%) \\
Transitional  & 2/89 (2.2$^{+2.8}_{-0.7}$\%) & 0/157 ($<$2.3\%; 95\% C.L.)   & 0/118 ($<$3.0\%; 95\% C.L.) \\
Evolved       & 2/89 (2.2$^{+2.8}_{-0.7}$\%) & 1/157 (0.6$^{+1.4}_{-0.2}$\%) & 0/118 ($<$3.0\%; 95\% C.L.) \\
\hline
\end{tabular}
\begin{flushleft}
Spectroscopic Accretion and IR Excess fractions shown above for K-type members of Sco-Cen only, with masses $\sim$0.7-1.3\msun.
\end{flushleft}
\end{table*}

\subsubsection{Infrared Excess Disk Fraction}\label{sec:irexcesses}

\begin{figure}
\begin{center}
\includegraphics[scale=0.45]{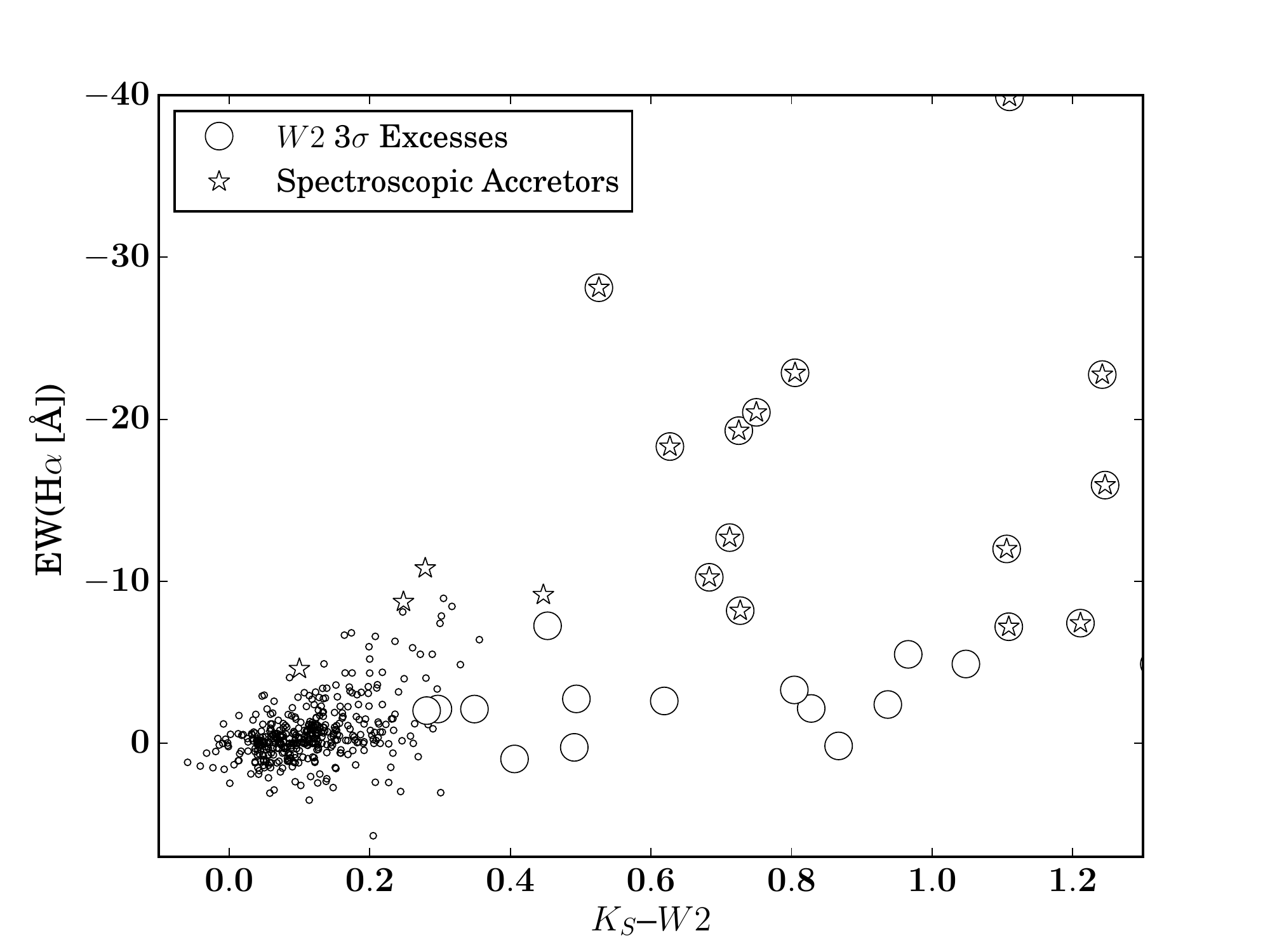}
\caption{$K_S$--$W2$ color versus EW(H$\alpha$) for Sco-Cen members.
  Members spectroscopically identified as accretors with the
  \protect\cite{barrado2003} criteria are shown as stars.  Members
  exhibiting a $K_S$--$W2$ color excess are shown as large circle.}
\label{fig:kw2_halpha}
\end{center}
\end{figure}

\begin{figure*}[]
\begin{center}
\includegraphics[scale=0.40]{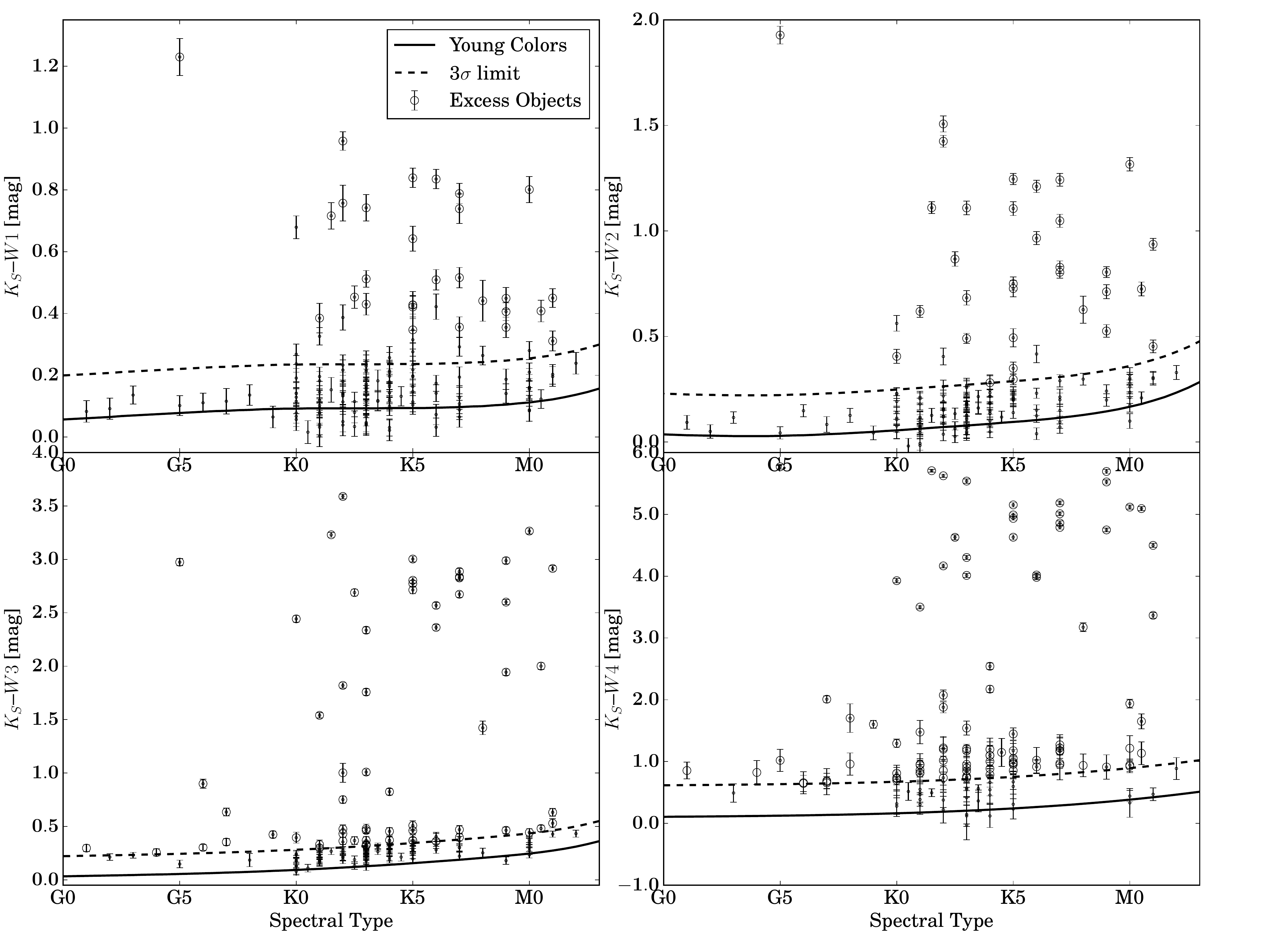}
\caption{Spectral type versus infrared colors $K_S$--$W1$,
  $K_S$--$W2$, $K_S$--$W3$, and $K_S$--$W4$ for Sco-Cen members.  The
  solid lines are the pre-MS photospheric colors from
  \protect\cite{pecaut2013}.  Members with color excesses above the
  3$\sigma$ dispersion of photospheric colors in that band and all
  bands at longer wavelength are identified as having excess emission
  from circumstellar disks.  Objects with color uncertainties greater
  than 0.25~mag are not shown.}
\label{fig:spt_kwise}
\end{center}
\end{figure*}

Infrared photometry can be used to identify the presence of
circumstellar matter with different wavelengths used to probe matter
at different temperatures.  To probe cooler, dusty debris around young
stars, 20$\mu$m and longer wavelengths are useful.  We examine Sco-Cen
members for excesses in $H$--$K_S$, $K_S$--$W1$, $K_S$--$W2$,
$K_S$--$W3$, and $K_S$--$W4$ colors, shown in \autoref{fig:spt_hk}
and \autoref{fig:spt_kwise}, in order to identify Sco-Cen members
exhibiting IR excesses which may indicate the presence of a disk and
allow us to infer its evolutionary state.  These stars are also
plotted against H$\alpha$ equivalent width (EW(H$\alpha$)) in
\autoref{fig:kw2_halpha}, demonstrating that accreting stars
identified with the \cite{barrado2003} EW(H$\alpha$) criteria also
typically exhibit a $W1$ band excess due to the presence of a warm
circumstellar gas disk.  We identify stars above the 3$\sigma$
dispersion in the young stellar locus, as defined in
\cite{pecaut2013}, as having an excess in that band.  We use 
3$\sigma$ as a conservative criterion, to avoid reporting excesses
with a small confidence level.  To avoid reporting false excesses 
due to scatter in the photometry, we only report an infrared excess 
if that object is also above the 3$\sigma$ dispersion in the young 
stellar locus for that band {\it and} all bands at longer wavelengths.

\begin{figure}
\begin{center}
\includegraphics[scale=0.45]{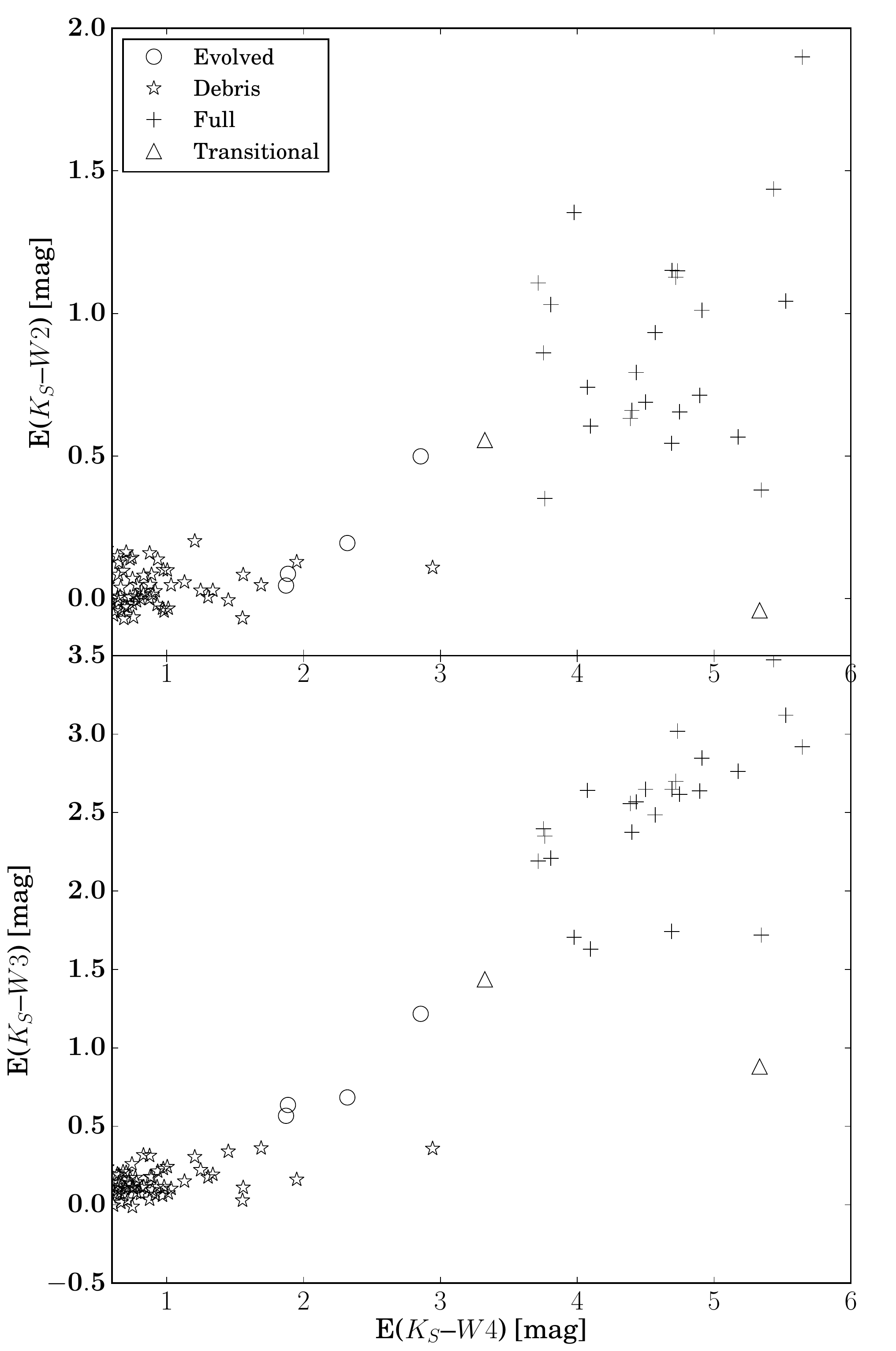}
\caption{Color excesses above the photosphere for stars in Sco-Cen
  exhibiting an infrared excess.}
\label{fig:diskclasses}
\end{center}
\end{figure}

Disks in young populations such as Sco-Cen may be found in a variety
of stages of evolution.  Here we attempt to classify the disks into
the disk classification scheme described by \citet{espaillat2012},
using the observational criteria described by \citet{luhman2012}.
Based on the boundary between the full disks and transitional, evolved
and debris disks for the stars classified in \citet{luhman2012}, we
identify ``full disks'' as those with E($K_S$--$W3$) $>$ 1.5 and
E($K_S$--$W4$) $>$ 3.2, ``transitional disks'' as those with
E($K_S$--$W4$) $>$ 3.2 but have E($K_S$--$W3$) $<$ 1.5.  We then
identify ``evolved disks'' as those with E($K_S$--$W4$) $>$ 3.2 and
E($K_S$--$W3$) $>$ 0.5 and ``debris disks'' as those with
E($K_S$--$W4$) $<$ 3.2 and E($K_S$--$W3$) $<$ 0.5.  Color excesses and
classifications are shown in \autoref{fig:diskclasses}.  Four stars
had excesses in $W1$, $W2$ or $W3$ but no reliable $W4$.  We classify
them as evolved or debris disks based on their lack of spectroscopic
signatures of accretion and their E($K_S$--$W2$) and E($K_S$--$W3$).

Disks classified using this method are listed in
\autoref{tbl:member_properties}.  Our sample contains a small number 
of G- and M-type members of Sco-Cen.  Therefore, in 
\autoref{tbl:excess_stats} we summarize infrared excess fractions 
for only K-type ($\sim$0.7-1.3~\msun) Sco-Cen members in the 2MASS 
$K_S$ band and the four WISE bands.

\section{Discussion}

\subsection{Which Ages are Reliable?}\label{sec:discuss_ages}

Depending on which isochronal ages we examine, we obtain different
mean subgroup ages.  However, given the {\teff}-dependent age trend
present in \autoref{fig:hrd_iso}, we see several reasons to
distrust ages from M-type stars.  The most obvious reason is that
evolutionary models have difficulty predicting the radii of
main-sequence M-type stars, systematically underestimating their radii
by $\sim$5-20\% (\citealt{torres2002,boyajian2012b}; see also
\citealt{kraus2011}).  \citet{hillenbrand2004} have compared
dynamically constrained masses with predictions from pre-MS
evolutionary models and found that the models systematically
underpredict the masses of stars 5\%-20\% for masses under
0.5~M$_{\odot}$.  On the other hand, they found that for masses above
1.2~M$_{\odot}$, dynamical masses and predicted masses from all models
are consistent.  As stated in \citet{hillenbrand2008}, the likely
source for the poor performance of models in the low-mass regime is
incomplete knowledge of opacity sources and the difficulty in modeling
convection.  One of the large sources of uncertainty particularly
applicable to modeling low-mass stars is the role magnetic fields play
in convection, as discussed in \citet{feiden2016}.  Strong magnetic 
fields can give rise to large star spots on the surface of the star, 
which will may cause the star to exhibit more than one surface 
temperature \citep{stauffer2003}.  \citet{somers2015b} have studied 
the effect of star spots on inferred
ages and masses in pre-main sequence stars, and find that spots will
tend to inflate the radii of young stars and shift them to a cooler
\teff.  The effect is that spots will make older stars appear to be
younger and less massive than implied by the evolutionary models.
\citet{somers2015b} offer age and mass correction factors which can be
used to estimate less biased, more accurate masses and ages from
published evolutionary tracks.  Given the problems with radii and mass
discrepancies from models as well as the potential influence of spots,
and that there are higher-mass F- and G-type isochronal ages
available, it is preferable to avoid the adopting ages from the K- and
M-type members.

What about kinematic ages?  Kinematic ``expansion'' or ``traceback''
ages are not dependent on stellar interior models and therefore offer
the prospect of nearly model-free ages.  \cite{song2012} argues that
kinematic expansion ages can function as model-independent age
benchmarks which can then be used to establish a model-indepedent age
scale.  Age indicators such as Li can then be used to establish
relative ages among different stellar populations.  This is an
interesting idea in principle, however there are a few major issues
with kinematic traceback ages.  

One major problem is that one must assume that the association was in 
a physically smaller configuration at some time in the past.  The 
data presented in section~\ref{sec:age_map} indicates substantial 
substructure in Sco-Cen.  This is consistent with work in other OB 
associations, for example, Cygnus OB2 also exhibits substantial 
substructure \citep{wright2014,wright2016}, and was never a compact, bound 
cluster.  The presences of substantial substructure invalidates the 
assumption that the members of these associations were in more 
compact configurations in the past, which prevents the determination 
of a meaningful kinematic age.

Another major issue is that the results from kinematic ages are 
very sensitive to the implementation.  A recent example 
is that in the TW Hydra Association (TWA).  A commonly quoted age for 
TWA is 8~Myr based on the kinematic traceback of \cite{delareza2006}.  
However, their kinematic traceback was based on a sample of only four 
stars, which was contaminated by TWA~19, a member of LCC 
\citep{mamajek2002}.  \cite{mamajek2005} calculated a kinematic 
expansion age for TWA using kinematic parallaxes and a vetted list of 
members and obtained a lower limit of $\sim$10~Myr on the expansion 
age at 95\% confidence, though the data were only weakly consistent 
with expansion.  More recently, \cite{weinberger2012} performed a 
kinematic traceback of TWA using a vetted list of members and 
trignometric parallaxes.  The \cite{weinberger2012} traceback result 
indicated the members were never in a significantly more compact 
configuration.  Another TWA study, \cite{ducourant2014}, independently 
obtained trignonmetric parallaxes for 13 stars, identify 31 as a 
co-moving association, 25 of which had radial velocity and 
trignometric parallax data.  This study obtained a traceback age for 
TWA of 7.5$\pm$0.7~Myr.  However, this result was based on a sample 
of 16 stars with converging motions, obtained after removing 9 stars 
which systematically drifted from the center of the association when 
traced back in time.  Similarly, an often-quoted age for Upper Sco is 
the 5~Myr expansion age derived from proper motion data by 
\cite{blaauw1978}.  However, it was demonstrated by \cite{brown1997a}, 
using simulations of expanding OB associations, that expansion ages 
inferred from proper motions alone all converged to $\sim$4~Myr, no 
matter the actual kinematic age.  A recent examination of the 
expansion age in Upper Sco by \cite{pecaut2012}, using radial velocity 
data, gave a lower limit of $\sim$10~Myr at 95\% confidence, though 
the data were consistent with no expansion.
Finally, we mention that the adopted kinematic expansion age of the
$\beta$ Pictoris moving group (BPMG) of 12~Myr, estimated by
\citet{song2003}, has been re-evaluated by \citet{mamajek2014}.  The
\citet{mamajek2014} study found the modern BPMG kinematic data was
only weakly indicitave of expansion, and that the age is only weakly
constrained by the kinematic data, giving a 95 confidence limit of
13-58 Myr.  We conclude that the there is no well-constrained
kinematic traceback age for either Sco-Cen or the groups used to
bracket its age (e.g., TWA, BPMG) that has withstood the scrutiny of
improved data, and that they simply do not yield useful age
constraints given the current precision of the available data.

Another relevant chronometric technique is the use of the lithium
depletion boundary (LDB) to determine the age of a stellar population.
By detecting the stellar \teff\, or luminosity above which all the
stars have exhibit Li depletion and comparing this with evolutionary
model predictions, one can obtain an age which is independent of
distance.  LDB ages have been calculated for several of the nearby,
young moving groups \citep{mentuch2008,binks2014} but the subgroups of
Sco-Cen do not yet have a reliable LDB age.  The results of
\cite{cargile2010} suggests that lithium depletion boundary ages and
modern nuclear main sequence turn-off ages are in agreement when
convective core overshooting is included in the models of high-mass
stars \citep[e.g.,][]{ekstrom2012}.  However, LDB ages are typically
much older than pre-MS contraction ages, and it has been suggested
that this problem is related to the radii discrepancy in M-type stars
\citep{yee2010,somers2015b}.

Recent discoveries of eclipsing binaries, particularly in Upper Scorpius
using the data obtained using the Kepler K2 mission, should help 
evolutionary models significantly by providing well-constrained radii 
and masses at these young ages.  Particularly, the recently published 
discoveries by \citet{david2015}, \citet{alonso2015}, \citet{lodieu2015},
and \citet{kraus2015}, will add a signifcant number of benchmark 
eclipsing binaries with tightly constrained, nearly model-independent 
parameters for objects in Upper Scorpius.  Masses and radii for the 
eclipsing binary UScoCTIO~5 from \citet{kraus2015}, when compared to 
the \citet{baraffe2015} evolutionary models, already provide some 
concordance with the older $\sim$10~Myr ages from the F- and G-type 
stars, though more theoretical work remains to be completed.

\subsection{How ``Coeval'' are the Sco-Cen Subgroups?}

Previous studies of Sco-Cen have attempted to quantify the observed
age spread in the subgroups.  In Upper Sco, \cite{preibisch2002},
adopting an age of 5~Myr, concluded that the age spread was
$\simless$1-2~Myr.  Their results account for the effects of binarity,
a distance spread, and observational uncertainties.  This is
consistent with the results of \citet{slesnick2006}, who similarly
constrained the age spread in the northern part of Upper Sco to be
less than $\pm$3~Myr (uniform distribution) using similar assumptions.
In their study of UCL and LCC, \citet{mamajek2002} have examined age
spreads in the older subgroups and have constrained the star formation
to have occurred over a time period of $\pm$3~Myr and $\pm$2~Myr
(1$\sigma$) for UCL and LCC, respectively.

Our age spreads are larger than those previously reported, with
1$\sigma$ age spreads of $\pm$7~Myr, $\pm$7~Myr, and $\pm$6~Myr for
US, UCL and LCC, respectively.  Our age spread of $\pm$7~Myr for Upper
Sco is much larger than the age spreads detected by
\citet{preibisch2002} and \citet{slesnick2008}.  However, the age map
in \autoref{fig:agemap} indicates that there is an age gradient
from the southeastern part of US to the northwest, with the
northwestern part being younger.  The \citet{slesnick2008} and
\citet{preibisch2002} low-mass samples were drawn from smaller regions
($\sim$150~deg$^2$ and $\sim$160~deg$^2$, respectively) than our
sample in Upper Sco (drawn from the entire $\sim$320~deg$^2$), which
could be responsible for the smaller detected age spreads.  However,
the likely reason our inferred age spreads in US are larger than
previous results is that we adopt a mean age of 10~Myr, twice as old
as the \citet{slesnick2008} or \citet{preibisch2002} studies.  A given
luminosity spread at a younger age corresponds to a smaller inferred
dispersion in ages than the same luminosity spread at an older age,
simply because the younger isochrones are spaced farther apart in
luminosity than those at older ages.

\citet{slesnick2008} suggests that spreads in H-R diagram positions
may not be an accurate proxy for spreads in age.  They demonstrate
this by comparing two nearly identical spectra for members in US, with
spectroscopic surface gravity indicators which are indicitive of
nearly identical surface gravity.  However, their H-R diagram
positions suggests their ages differ by more than 10~Myr!
\citet{jeffries2011} use constraints on the disk lifetime to show that
the age spread in the Orion Nebula Cluster must be less than 0.14~dex
in $\log$(Age), though the age spreads inferred from the H-R diagram
show a 0.4~dex dispersion in $\log$(Age).  These results suggest that
scatter in the H-R diagram may not be a reliable indicator of age
spreads. However, the H-R diagram remains the best observational 
indicator available at the present time for revealing any intrinsic 
age spreads.  See \citet{soderblom2014} for a more detailed and 
complete discussion regarding age spreads.

As stated previously, our age map does indicate there is a clear age
gradient in US as it merges into UCL.  The southern part of LCC is
also noticably younger than other parts of LCC, confirming the
suggestion first raised in \citet{preibisch2008}.  We note that there
are no main-sequence turnoff stars in northern LCC, which, considering
the younger age of southern LCC, accounts for the turn-off age of LCC
being much younger than UCL.  The star formation history of Sco-Cen, 
as inferred from the H-R diagram positions, appears to be more complex 
than previously treated.  The spatial distribution of ages is 
suggestive that the current division of three distinct, coeval 
subgroups is overly simplistic and a separation into smaller units may 
be warranted.  We avoid speculating further on scenarios of triggered
star formation here, leaving that discussion to a future study.

\subsection{Circumstellar Disk Census}

\begin{figure}
\begin{center}
\includegraphics[scale=0.45]{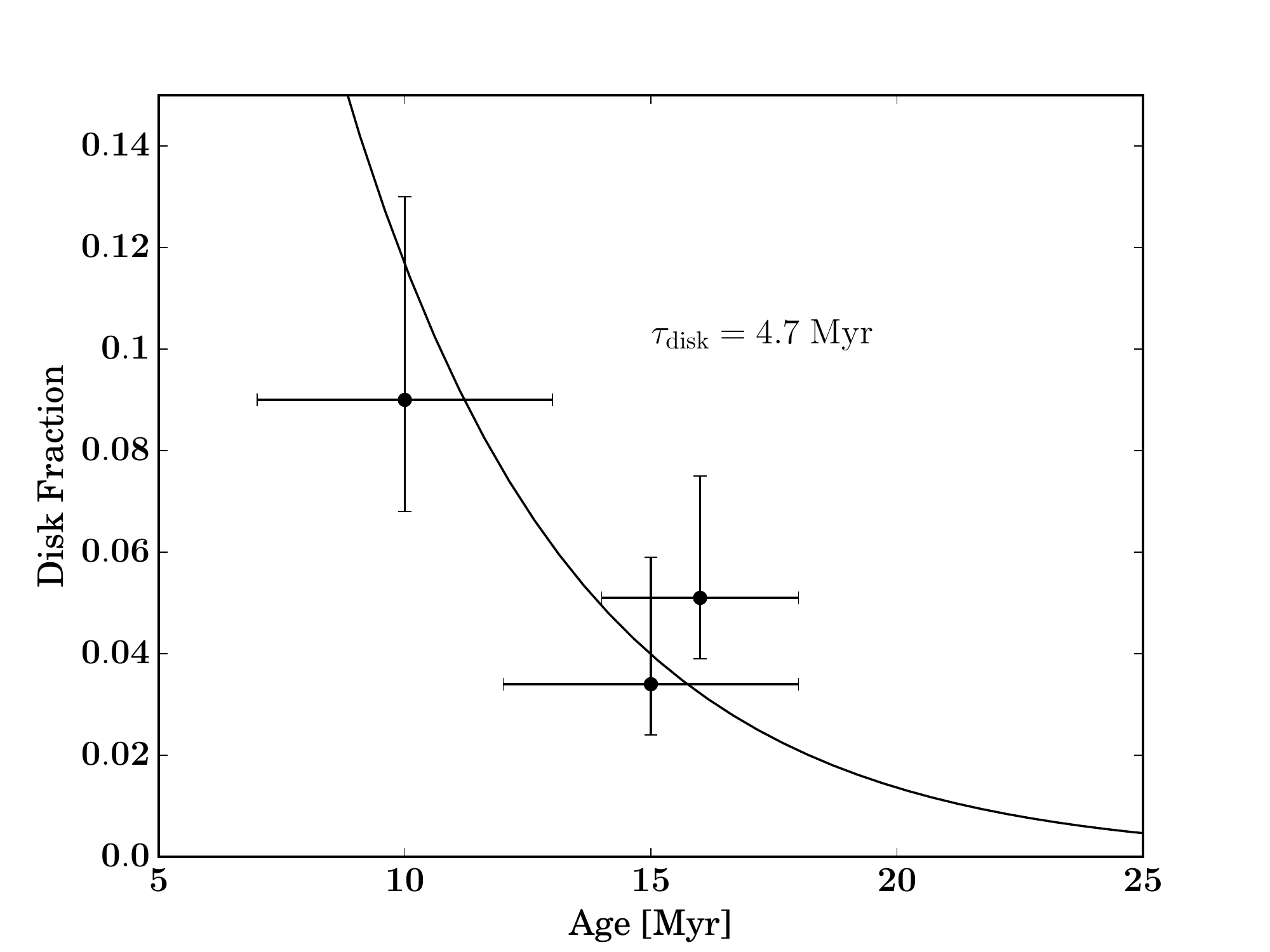}
\caption{Mean subgroup age versus disk fraction (``Full Disk''; see
  \autoref{tbl:excess_stats}) for K-type stars, $\sim$0.7-1.3\msun, 
  from the three subgroups of Sco-Cen.  The best-fit exponential 
  decay curvey $f_{disk}=e^{-t/\tau_{disk}}$ has 
  $\tau_{disk}$=4.7~Myr.}
\label{fig:disk_decay}
\end{center}
\end{figure}

Observations of young clusters and associations of ages from $\sim$1
to $>$100~Myr have given strong indication that the protoplanetary
disk dispersal timescale is very short, with an e-folding time of
$\sim$ 2.5~Myr \citep{mamajek2009,fedele2010}.  This is qualitatively 
consistent with what we seen in Sco-Cen: 9\% of Upper Sco K-type stars 
host an optically thick protoplanetary disk (``Full Disk'' in the
\citealt{espaillat2012} nomenclature) at an age of $\sim$10~Myr,
whereas $\simeq$4\% of the K-type stars in the older subgroups UCL and
LCC host a full disk (\autoref{tbl:excess_stats}).  However, the
e-folding time of 2.5~Myr was estimated adopting an Upper Sco age of
5~Myr, along with many other young clusters.  \citet{naylor2009} has
argued that pre-MS ages systematically underestimate cluster ages, and
that ages based on high-mass stars, typically double the ages estimated
from the low-mass stars, are more likely to be correct \citep{bell2013}.
Using the ages we adopt for the Sco-Cen subgroups, what disk dispersal 
timescale does this imply?  We plot the primordial disk fractions 
(``Full Disk'' from \autoref{tbl:excess_stats}) as a function of age in
\autoref{fig:disk_decay} and, following \citet{mamajek2009}, fit an
exponential decay curve to the data ($f_{disk}=e^{-t/\tau_{disk}}$).
We obtain a mean protoplanetary disk e-folding timescale of 4.7~Myr 
for K-type stars ($\sim$0.7-1.3\msun).  This is $\sim$2~Myr longer 
than the timescale estimated in \citet{mamajek2009}, and would imply 
a longer timescale available for planet formation, but consistent with 
the findings of \cite{bell2013}.

Finally, we note that our K-type disk fraction is larger than what is
observed for the higher mass-stars in the same subgroups
\citep{mamajek2002,pecaut2012,carpenter2009,luhman2012}.  This is
consistent with the mass-dependent trend first identified by
\citet{carpenter2006} and further confirmed by \citet{luhman2012} in
Upper Sco, as well as the results of \citet{hernandez2007} from the
younger $\sigma$~Ori association at $\sim$6 Myr \citep{bell2013}.  
\citet{ribas2015} have summarized these mass-dependent disk fraction 
trends using the nearby young stellar associations within $\sim$500 pc, 
and have come to a similar conclusion -- that the disk dispersal time 
depends on stellar mass, with the low-mass stars retaining their disks 
longer.  For Sco-Cen, however, we still have very poor statistics for 
M-type stars in the older subgroups, which highlights the need for 
future surveys to push the membership census to cooler spectral types 
in UCL and LCC \citep[e.g.,][]{murphy2015}.

\section{Conclusions}

We can summarize the findings from our survey as follows:

\begin{enumerate}

\item We have performed a survey for new, low-mass K- and M-type
  members of all three subgroups of Sco-Cen.  Using Li, X-ray, and 
  proper motion data, we identify 156 new pre-MS members of Sco-Cen.

\item Using our newly identified members together with previously
  known members of Sco-Cen, we utilize H$\alpha$ as an accretion
  diagnostic and identify stars with H$\alpha$ emission levels
  consistent with accretion.  We estimate a spectroscopic accretion
  disk fraction of 7.1$^{+3.9}_{-1.9}$\%, 3.4$^{+2.2}_{-1.0}$\% and
  3.4$^{+2.5}_{-1.0}$\% for solar analog pre-MS K-type stars
  ($\sim$0.7-1.3~\msun) in US, UCL and LCC, respectively, consistent
  with a protoplanetary disk decay e-folding timescale of 
  $\sim$4.7~Myr, or half-life of $\sim$3.3~Myr.

\item Similar to previous results in other star-forming regions
  \citep[e.g.,][]{hillenbrand1997,hillenbrand2008,bell2013}, we 
  observe a \teff-dependent age trend in all three subgroups of 
  Sco-Cen, for all sets of evolutionary tracks.  

\item We adopt median ages of 10$\pm$3~Myr, 16$\pm$2~Myr and
  15$\pm$3~Myr for US, UCL and LCC, respectively, when considering the
  revised nuclear ages as well as the pre-MS contraction ages from the
  F- and G-type stars.

\item We obtain estimates for the intrinsic age spread in each
  subgroup through a grid of Monte Carlo simulations which take into
  account binarity, spots, and observational uncertainties.  Assuming the
  median ages obtained above, and modeling the age distribution as a
  gaussian, we find that 68\% of the star formation in US, UCL, and
  LCC occured over timescales of $\pm$7~Myr, $\pm$7~Myr, and
  $\pm$6~Myr, respectively.  Thus when adopting an age of $\sim$10~Myr
  for Upper Sco, we detect an intrinsic age spread of $\pm$7~Myr
  (1$\sigma$).

\item Using members from our X-ray sample as well as F- and G-type
  members of Sco-Cen, we create an age map of the Sco-Cen complex.  We
  find that the star-formation histories of the UCL and LCC, the older
  subgroups, are indicitative of substructure and are not consistent
  with a simple triggered star-formation scenario.  The groups are not
  each monolithic episodes of star formation, but likely an ensemble of
  small subgroups, exhibiting significant substructure.

\end{enumerate}

\section*{Acknowledgements}

We would like to thank the anonymous referee, whose comments and
suggestions improved the quality of this work.
This work has been supported by funds from the School of Arts and
Sciences at the University of Rochester and NSF grants AST-1008908 and
AST-1313029.  EEM acknowledges support from the NASA Nexus for 
Exoplanet System Science program (NExSS).
We thank Fred Walter for the use of his SMARTS
RC-spectrograph pipeline for reducing the data obtained at the SMARTS
1.5m telescope at Cerro Tololo, Chile, as well as Jose Velasquez and
Manuel Hernandez for their help and advice at the telescope.  This
research was made possible through the use of the AAVSO Photometric
All-Sky Survey (APASS), funded by the Robert Martin Ayers Sciences
Fund.  This publication makes use of data products from the Two Micron
All Sky Survey, which is a joint project of the University of
Massachusetts and the Infrared Processing and Analysis
Center/California Institute of Technology, funded by the National
Aeronautics and Space Administration and the National Science
Foundation.  This publication makes use of data products from the
Wide-field Infrared Survey Explorer, which is a joint project of the
University of California, Los Angeles, and the Jet Propulsion
Laboratory/California Institute of Technology, funded by the National
Aeronautics and Space Administration.



\bibliographystyle{mnras}
\bibliography{ms} 




\appendix

\bsp	
\label{lastpage}
\end{document}